\documentclass[pop,english,amssymb,amsmath]{revtex4-1}
\usepackage[T1]{fontenc}
\usepackage[latin9]{inputenc}
\usepackage{xcolor}
\usepackage{float}
\usepackage{textcomp}
\usepackage{amstext}
\usepackage{graphicx}
\usepackage{subcaption}
\usepackage{adjustbox}
\PassOptionsToPackage{normalem}{ulem}
\usepackage{ulem}
\usepackage{babel}
\usepackage{hyperref}

\begin{document}

\title{Onset of internal transport barriers in tokamaks}

\author{L. A. Osorio}
\affiliation{Instituto de F{\'i}sica, Universidade de S{\~a}o Paulo, S{\~a}o Paulo,
SP 05315-970, Brazil}

\author{M. Roberto}
\affiliation{Departamento de F{\'i}sica, Instituto Tecnol{\'o}gico da Aeron{\'a}utica,
S{\~a}o Jos{\'e} dos Campos, SP 1228-900, Brazil}
\email{Author to whom correspondence may be addressed: r.marisa@gmail.com}

\author{I. L. Caldas}
\affiliation{Instituto de F{\'i}sica, Universidade de S{\~a}o Paulo, S{\~a}o Paulo, SP 05315-970 - Brazil}

\author{R. L. Viana}
\affiliation{Departamento de F{\'i}sica, Universidade Federal do Paran{\'a}, Curitiba, PR 81531-990, Brazil}

\author{Y. Elskens}
\affiliation{Aix-Marseille university, UMR 7345 CNRS, PIIM, campus Saint-J\'er\^ome, case 322, av.\ esc.\ Normandie-Niemen, 52, FR-13397 Marseille cedex 13, France}


\begin{abstract}
Barriers have been identified in magnetically confined plasmas reducing the particle transport and improving the confinement. One of them, the primary shearless barriers are associated to extrema of non-monotonic plasma profiles. Previously, we identified these barriers in a model described by a map that allows the integration of charged particles motion in drift waves for a long time scale. In this work, we show how the existence of these  robust barriers depends on the fluctuation amplitude and on the electric shear. Moreover, we also find control parameter intervals for which these primary barriers onset and break-up are recurrent. Another noticeable feature, in these transitions, is the appearance of a layer of particle trajectory stickiness after the shearless barrier break-up or before its onset. Besides the mentioned primary barriers, we also observe sequences of secondary shearless barriers, not reported before, created and destroyed by a sequence of bifurcations as the main control parameters, the fluctuation amplitude and electric shear, are varied. Furthermore, in these bifurcations, we also find hitherto unknown double and triple secondary shearless barriers which constitute a noticeable obstacle to the chaotic transport \footnote{This article has been accepted by the Journal Physics of Plasmas. It can be found at: \url{https://doi.org/10.1063/5.0056428}}. 
\end{abstract}

\maketitle

\section{\label{I}Introduction}

One of the questions of paramount importance in the quest for magnetically confined fusion plasmas is the understanding and control of radial particle transport \cite{horton99,wolf,hazeltine}. In particular, in order to build a future tokamak-based fusion reactor, it is mandatory to improve the energy confinement by reducing particle transport to acceptable levels \cite{bookHB}.\\

Auxiliary heating applied to tokamak discharges often leads to low confinement (L-mode) plasma regimes, for which there is an enhanced radial cross-field transport due to a high level of turbulence \cite{goldston,yushmanov}. By combining neutral beam heating and a divertor, it was possible to obtain a high confinement regime (H-mode) for tokamak plasmas, with reduced transport fluxes \cite{wagner,connor,garbet}.\\

In the latter, there is a relatively high pedestal pressure profile in the plasma column, and a large pressure gradient at the plasma edge. This increased gradient is related to a local reduction of the turbulence levels due to ${\bf E}\times{\bf B}$ shear \cite{burrell}. The edge transport barriers (ETBs) which are present in the H-mode can also exist inside the plasma core and are called generally internal transport barriers (ITBs), which are regions of reduced radial particle transport \cite{wolf}. On both cases, the barriers are characterized by steep pressure gradients, however.\\

Another type of internal transport barrier has been recently proposed, the so-called shearless transport barriers (STBs), for which the pressure gradients are not necessarily high as in ITBs. The basic mechanism underlying the STBs is the existence of non-monotonic equilibrium radial profiles inside the tokamak. These profiles can be created by modifications of the plasma current profile and/or the application of radial electric fields \cite{shearless}.\\

An example of vanishing magnetic shear occurs for tokamak plasmas with a non-monotonic safety factor $q(r)$ profile, so that there can be one or more radial positions which are extremum points of $q(r)$ \cite{levinton, strait,mazzucato}. Particle transport is reduced for this type of non-monotonic discharges. At these points, a shearless toroidal magnetic surface is formed \cite{morrison}. Given an external perturbation with modes resonant with the magnetic surfaces in both sides of the shearless torus, twin resonant island chains are formed therein, which produce a local region of chaotic magnetic field lines attached to the islands' boundaries \cite{diego1,diego2}.\\

In a first approximation, plasma particles follow magnetic field lines which lie on magnetic surfaces, in such a way that there is no cross field transport. Within this picture, the tori surrounding the shearless curve act as dikes preventing chaotic transport. If we consider higher-order effects like finite Larmor-radius and collisions, there would be transport even in the presence of magnetic surfaces.\\

As the perturbation intensity increases, the width of these chaotic layers is also increased, engulfing other tori between the islands and the shearless torus. Increasing further the perturbation, all the tori can be destroyed and, even after the shearless torus is broken, it still acts as a transport barrier. This occurs because of a dynamical effect called stickiness, which makes a chaotic trajectory to wander erratically in the vicinity of the torus remnant, yielding trajectories with large escape times, so reducing effectively the transport fluxes \cite{szezech}.\\

There are other sources of vanishing shear that can be related to the formation of shearless transport barriers. The introduction of a bias electrode in the plasma column produces a radial electric field that improves plasma confinement in tokamaks \cite{nascimento}. In particular, a decrease was observed in the levels of the low-frequency component of the fluctuating floating potential as well as the turbulent-driven particle flux \cite{vanoost}. The production of shearless transport barriers due to a non-monotonic radial electric field produced by polarizing the tokamak vessel has been numerically investigated by considering a drift-kinetic model for particle transport driven by drift waves \cite{marcus,horton98, marcus2}.\\

A further physical mechanism which can also trigger the formation of shearless transport barriers is the presence of non-monotonic plasma toroidal velocity profiles \cite{ferro}. The existence of such barriers has been observed in the Texas Helimak, where a set of diagnostic probes is mounted to measure plasma flow velocity in various points, so that velocity shears can be detected in the discharges \cite{gentle,toufen}.\\

We have applied the model introduced in \cite{horton98} to predict numerically the existence of STBs in tokamak discharges with non-monotonic plasma profiles \cite{rosalem1,miskane,marcus}. In this paper, our main goal is to examine in details the onset of shearless transport barriers in tokamaks associated to one of those described sources of non-monotonic plasma profiles, namely, the radial electric field. The common framework to deal with these physical mechanisms is the drift-kinetic model of Ref. \cite{horton98}, where we describe guiding center motion of plasma particles under ${\bf E}\times{\bf B}$ drift with contributions of the parallel velocity $v_\parallel$ as well. The equations of motion are numerically integrated and Poincar\'e maps are obtained for discrete times.\\

Since the ensuing dynamical model is non-integrable, we typically observe chaotic motion related to the resonant boundaries. From the numerically obtained Poincar\'e maps, we are able to detect the formation of shearless transport barriers. We call these barriers primary because they can be associated to the non-monotonic equilibrium profile and the corresponding extrema of the rotation number profile.\\

Using the drift-kinetic model, we investigate the behavior of the primary shear transport barriers when model parameters are changed, particularly how they depend on the amplitude of electrostatic fluctuations and the extremum of radial electric field. We show that the barriers break up and resurge, recurrently, as those parameters are varied. In these transitions, after the shearless barrier break-up or before its onset, we also find the appearance of a layer of particle trajectory stickiness, which is a obstruction to chaotic transport.\\

Besides the mentioned primary barriers, we also  find secondary shearless
barriers recurrently created and destroyed by a sequence of bifurcations that creates new extrema in the rotation number profile. Furthermore, in these bifurcations, we also find hitherto unknown double and triple secondary shearless barriers which also constitute a noticeable obstacle to the chaotic transport.\\

The paper is organized as follows: Section \ref{sec:Drift_Model} introduces the drift-kinetic model and the non-monotonic plasma profiles used in numerical simulations. Section \ref{sec:Transport-barriers} discusses the influence of the fluctuation amplitude on the shearless transport barriers. The formation of effective barriers due to stickiness effect, after the shearless curve has been destroyed, is discussed in Section \ref{sec:Persistent-Barriers}. The influence of the radial electric field profile on the formation of shearless barriers is the subject of Section \ref{sec:fixed phi2}. The last Section is devoted to our Conclusions.

\section{Drift Wave model\label{sec:Drift_Model}}

The chosen drift wave model introduces the basic equations of motion to describe the trajectories of particles along magnetic field lines and electrical drifts. We consider an electrostatic equilibrium field in the radial direction and drift waves that propagate in the poloidal and toroidal directions. These drift waves arise from non-uniformity at the plasma edge and are analyzed in the toroidal section of magnetic confinement in the tokamak \cite{horton98,miskane,horton99,rosalem1,rosalem2,severo}. Fluctuations in electrostatic potential are written as a function of amplitude, and in spatial and temporal modes \cite{rosalem1,miskane}.\\

The trajectories of the particles are described by the movement of the guiding center

\begin{equation}
\frac{d\mathbf{x}}{dt}=v_{\parallel}\frac{\mathbf{B}}{B}+\frac{\mathbf{E}\times\mathbf{B}}{B^{2}}\label{eq:motion}
\end{equation}

\noindent We consider, initially, that the components of this equation are given in local polar coordinates $\mathbf{x}=(r,\theta,\varphi)$.\\

This plasma configuration corresponds to a cylindrical approximation of a toroidal section of a tokamak of high aspect ratio ($a/R\simeq0.3$), where $a$ and $R$ are, respectively, the minor and major plasma radius. For the field components, we assume a magnetic configuration with $B\approx B_{\varphi}\gg B_{\theta}$ and the safety factor given by  $q(r)=\frac{rB_{\varphi}}{RB_{\theta}}$. The electric field consists of an equilibrium part with intensity given by  $E_{r}$, defined by a radial profile, and a floating part, so that  $\tilde{\mathbf{E}}=-\nabla\tilde{\phi}$. Therefore, this model allows to investigate the simultaneous influence of the electric shear, due to the spatial variations of the electric field, and the magnetic shear, when considering the spatial variations of the safety factor, in the chaotic transport at the plasma edge.\\

We follow the procedure adopted in reference \cite{horton98}, writing the equations in the action variable $I\equiv(r/a)^{2}$ and angle variable $\psi\equiv M\theta-L\varphi$. Poloidal and toroidal spatial modes are defined, respectively, by the wave numbers $M$ and $L$. In this way we assume coherent oscillations, where $\psi$ represents a helical angle defined by dominant modes. When considering these new variable components of the guiding center, we obtain

\begin{eqnarray}
\frac{dI}{dt} & = & 2M\underset{n}{\sum}\phi_{n}\sin(\psi-n\omega_{0}t\text{+}\alpha_{n})\label{eq:acao}\\
\frac{d\psi}{dt} & = & \frac{a }{R}v_{\parallel}(I)\frac{[M-Lq(I)]}{q(I)}-\frac{M}{\sqrt{I}}E_{r}(I)\label{eq:angulo}
\end{eqnarray}

\noindent where a normalization with the characteristic magnitude scales $a$, $B$ and $E_0$ was done. Note that all the plasma profiles are defined as functions of the action variable. For the floating potential, we use a finite mode drift wave spectrum,

\begin{equation}
\tilde{\phi}(\mathbf{r},t)=\sum\limits _{n}\phi_{n}\cos(\psi-n\omega_{0}t+\alpha_{n})\label{eq:Fluctuating_potential}
\end{equation}

\noindent where ${\phi_{n}}$ is the perturbed amplitude, $\omega_{0}$ is the  lowest angular frequency of the drift wave spectrum and $\alpha_{n}$ are constant phases that do not affect the resonant conditions. Temporal modes are defined by the wave numbers $n$. Thus, we can assume one or more drift waves that describe fluctuations in the electrostatic potential of the model.\\

In the next sections, for numerical applications, we introduce typical parallel speed and radial electric field profiles experimentally measured in the TCABR tokamak \cite{severo}. We also assume a monotonic safety factor profile commonly observed in tokamaks \cite{bookHB}. The resonance conditions are determined by the combination of the safety factor, parallel plasma velocity and electric field profiles. Thus, taking $\frac{d}{dt}(\psi-n\omega_{0}t-\alpha_{n})=0$, we obtain the primary resonance condition. Furthermore, we also show the Poincar{\'e} maps by integrating the equations (\ref{eq:acao}) and (\ref{eq:angulo}) for various initial conditions chosen to clearly represent the most important islands, invariant lines and the chaotic region. The intersections of the integrated trajectories are selected at the toroidal section corresponding to instants $t_{j}=j\,2\pi/\omega_{0}$
$(j=0,1,2,...)$. In Poincar{\'e} maps, the minor plasma radius lies at $I = 1.0$, but we consider $I$ up to $1.4$ to investigate the particle transport to the chamber wall.

\section{Influence of the fluctuation amplitude on the shearless transport barriers\label{sec:Transport-barriers}}

According to equation (\ref{eq:acao}), for null perturbing amplitudes, $\phi_n = 0$, the system is integrable, it means that each trajectory is periodic or quasi-periodic and stays in invariant lines in the Poincar{\'e} map, with the initial action $I_0$ $=$ constant. When $\phi_n\ne 0$, we have chaotic trajectories and regular invariant lines and we can use the rotation number, defined as $\Omega=\lim_{i\to\infty}(\psi_{i}-\psi_{0})/i$,
where $\psi_{i}$ refers to the $i$-th section, to analyze the behavior of the invariant lines in the phase space. The rotation number profile is determined for initial conditions with a fixed angle $\psi_0$ for several values of action variable $I$. Although other choices of $\psi_0$ would result in  different profiles, the identified islands and other invariants would be essentially the same. This happens because each invariant and island depend on both variables $\psi$ and $I$, but are labeled by a unique identifying rotation number. If  $d\Omega/dI=0$, the profile has an extremum at $(I,\psi_{0})$. This point is part of a shearless invariant curve, which acts as a barrier separating the particle orbits in the phase space and reducing the particle transport. These are the primary shearless invariant curves if their origin can be associated to the rotation number extrema obtained from the equilibrium plasma profiles. However, in this work we also present secondary shearless invariant curves that are due to bifurcations that create new rotation number extrema for varying control parameters.\\

In order to solve our numerical model, we use the parallel velocity and radial electric field profiles, with $a=0.18$m, $B=1.1$T and $E_0=4.6$kV/m, as given in reference \cite{marcus}, which are similar to those observed in TCABR tokamak \cite{severo}. The velocity profile is given by equation (\ref{eq:velocity_profile}), which fits with experimental TCABR data points as described in reference \cite{marcus}. The equilibrium radial field $E_{r}$ is written in equation (\ref{eq:electric_profile}),
with $\alpha=-0.563$, $\beta=1.250$ and $\gamma=-1.304$.
Furthermore, we assume the safety factor profile as equation (\ref{eq:safety_factor_profile}), a common approximation for large aspect ratio tokamaks.\\

\begin{equation}
    v_{\parallel}(r)=-1.43+2.82\tanh\left(20.3\frac{r}{a}-16.42\right) \label{eq:velocity_profile}
\end{equation}
\begin{equation}
    E_{r}(r)=3\alpha\left(\frac{r}{a}\right)^{2}+2\beta\frac{r}{a}+\gamma \label{eq:electric_profile}
\end{equation}
\begin{equation}
    q(r)=1.0+3.0\left(\frac{r}{a}\right)^{2} \label{eq:safety_factor_profile}
\end{equation}

Applying the resonant condition for the chosen radial profiles, we verify that the resonant modes are $n=3$ and $n=4$.  We select from the spectrum analysis a frequency around $62.1$kHz, which gives $\omega_{0}=2.673$. The perturbing electric potential amplitudes $\phi_{n}$ are normalized by $aE_{0}$. We consider spatial wave numbers $M = 16$ and $L = 4$, chosen as typical numbers in the tokamak wave spectrum at plasma edge \cite{horton98}.\\

\begin{figure}[H]
  \centering
  	\begin{subfigure}[b]{0.35\linewidth}
	\includegraphics[width=\linewidth]{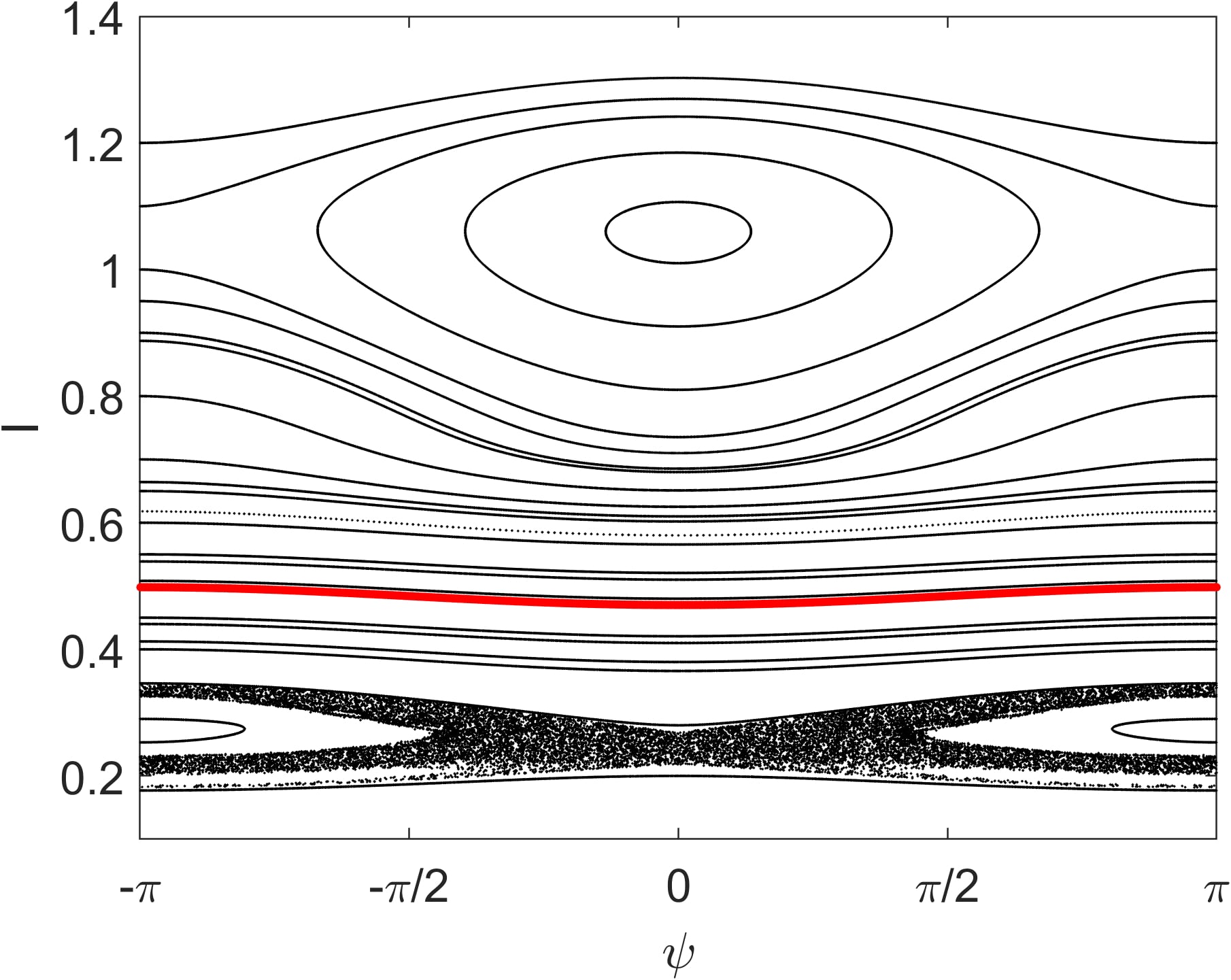}
	\caption{}
	\end{subfigure}	
    \begin{subfigure}[b]{0.35\linewidth}
	\includegraphics[width=\linewidth]{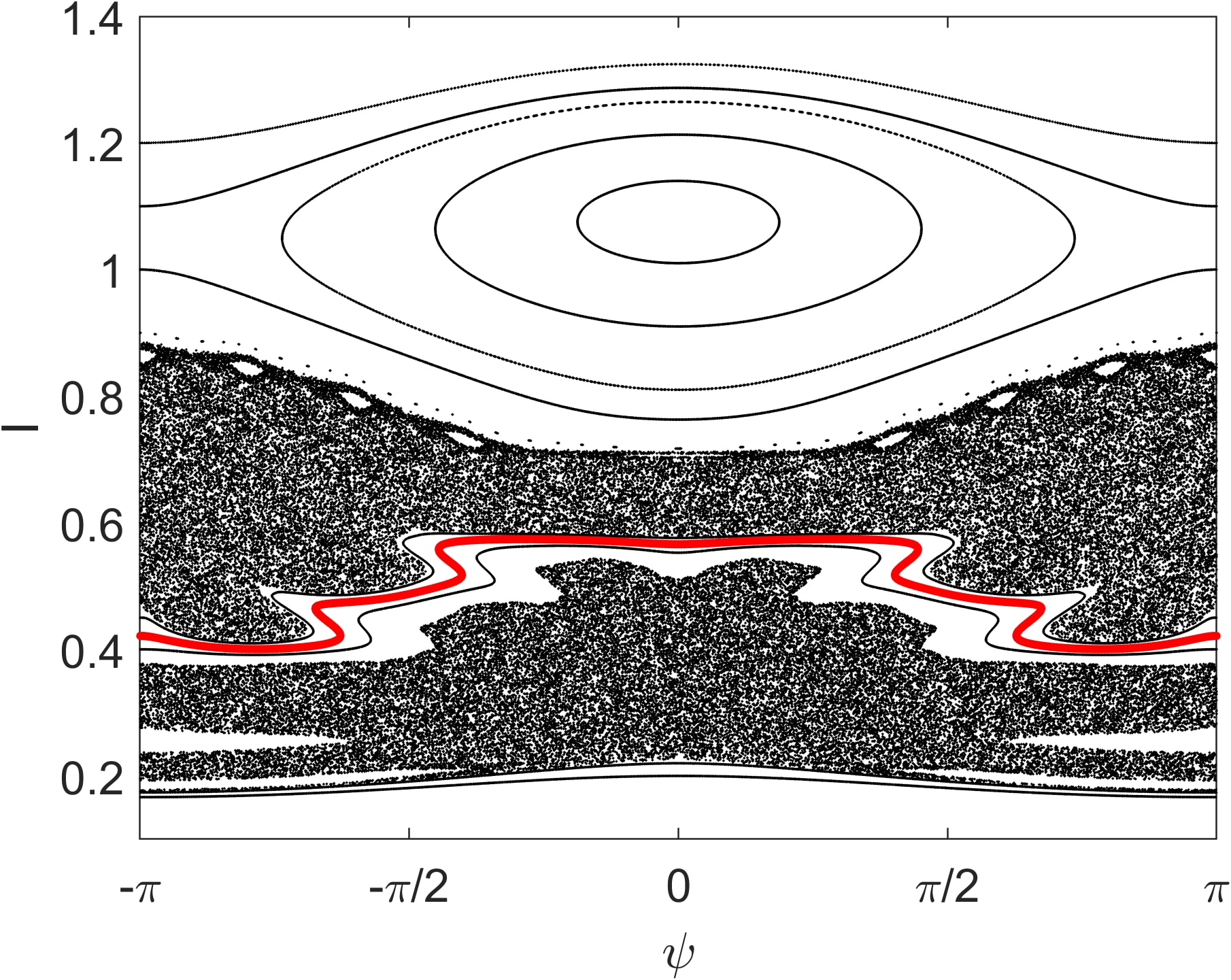}
	\caption{}
	\end{subfigure}	
	\begin{subfigure}[b]{0.35\linewidth}
	\includegraphics[width=\linewidth]{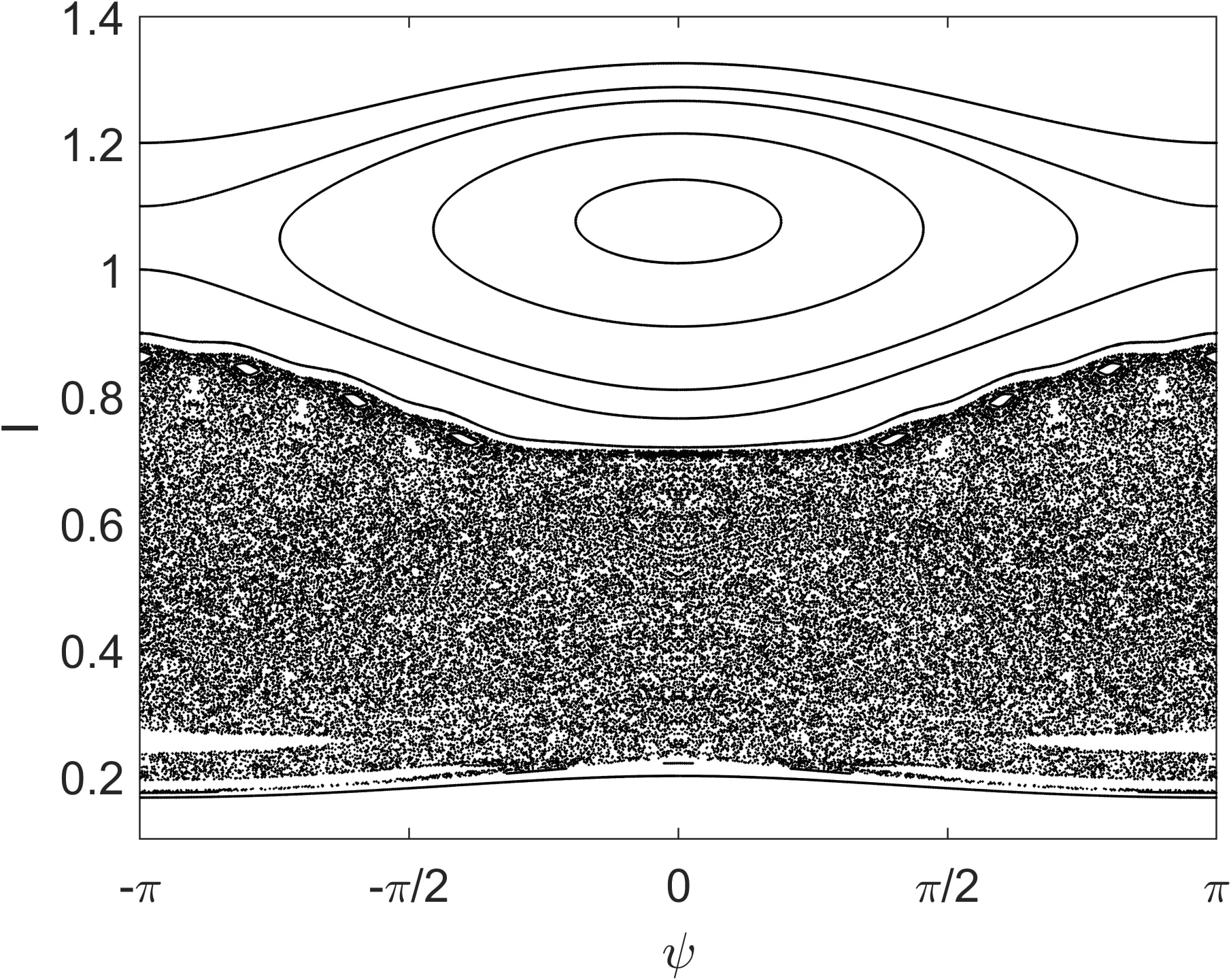}
	\caption{}
	\end{subfigure}	
    \begin{subfigure}[b]{0.35\linewidth}
	\includegraphics[width=\linewidth]{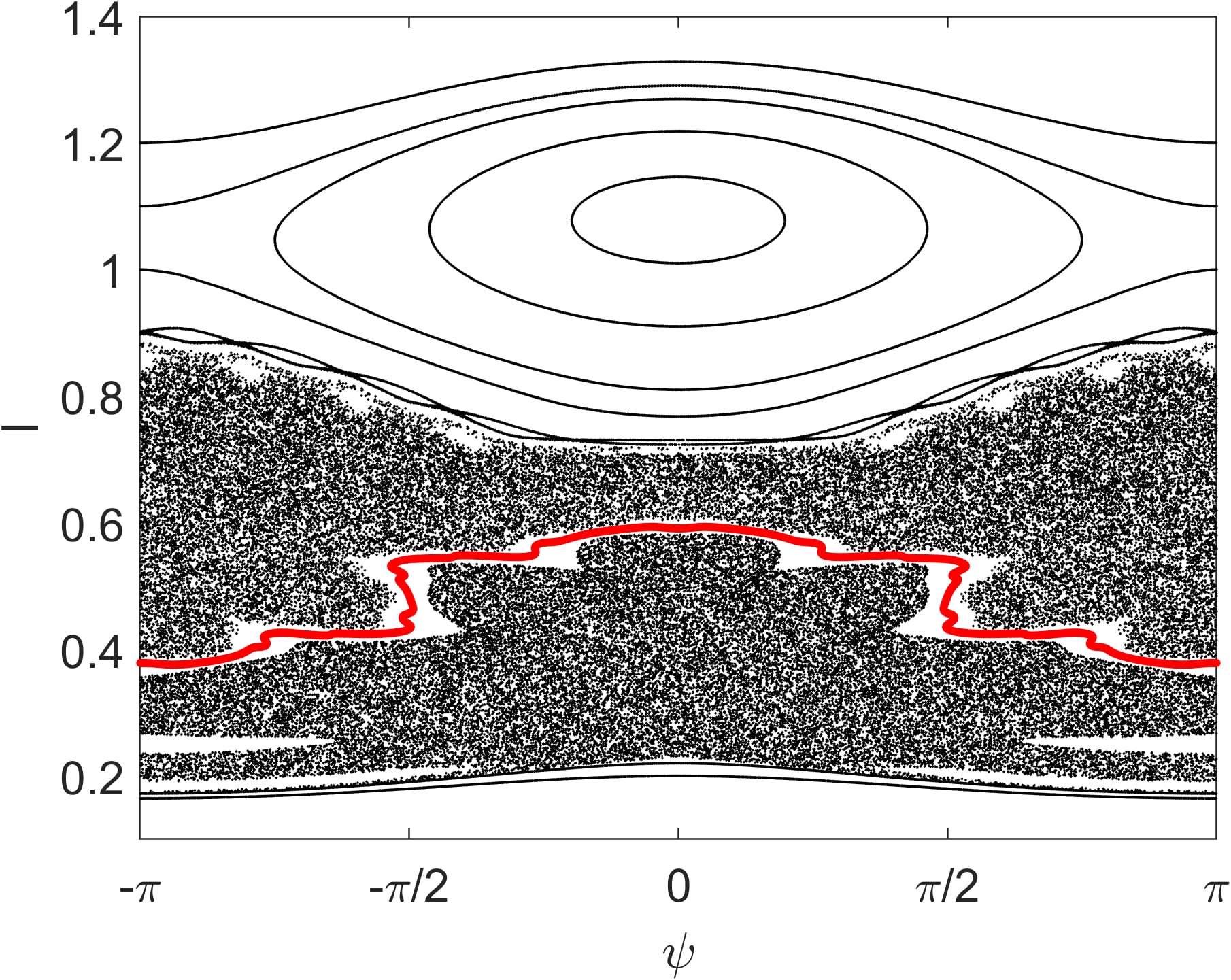}
	\caption{}
    \end{subfigure}
 \caption{Poincar{\'e} maps for (a) $\phi_{2} = 0$, (b) $\phi_{2} = 1.5\times10^{-3}$, (c) $1.6\times10^{-3}$ and (d) $1.8\times10^{-3}$. The shearless curve, marked in red, disappears for $\phi_{2}=1.6\times10^{-3}$ and resurges for $\phi_{2}=1.8\times10^{-3}$.}
  \label{fig:1}
\end{figure}

In order to analyze the influence of the fluctuation amplitudes on the shearless barriers, we keep fixed the perturbation amplitudes    $\phi_{3}=1.0\times10^{-3}$  and    $\phi_{4} = 0.12\times10^{-3}$ and vary the non-resonant mode $\phi_{2}$. We keep constant all the parameters of the equilibrium profiles. The objective is to investigate the recurrent break-up and onset of the barriers as the fluctuation amplitude is varied. Following that, Figure \ref{fig:1} shows the Poincar{\'e} maps for  $\phi_{2}= 0, 1.5\times 10^{-3}, 1.6\times 10^{-3}$ and $1.8\times 10^{-3}$. The barrier already exists for a null $n = 2$ amplitude, breaks up for the perturbing amplitude $\phi_{2}=1.6\times10^{-3}$ and resurges for $\phi_{2}=1.8\times10^{-3}$. So we see that the existence of the barrier is sensitive to small variations in the fluctuation amplitude, a characteristic that will be explored in this section.\\

\begin{figure}[h!]
  \centering
  	\begin{subfigure}[b]{0.35\linewidth}
	\includegraphics[width=\linewidth]{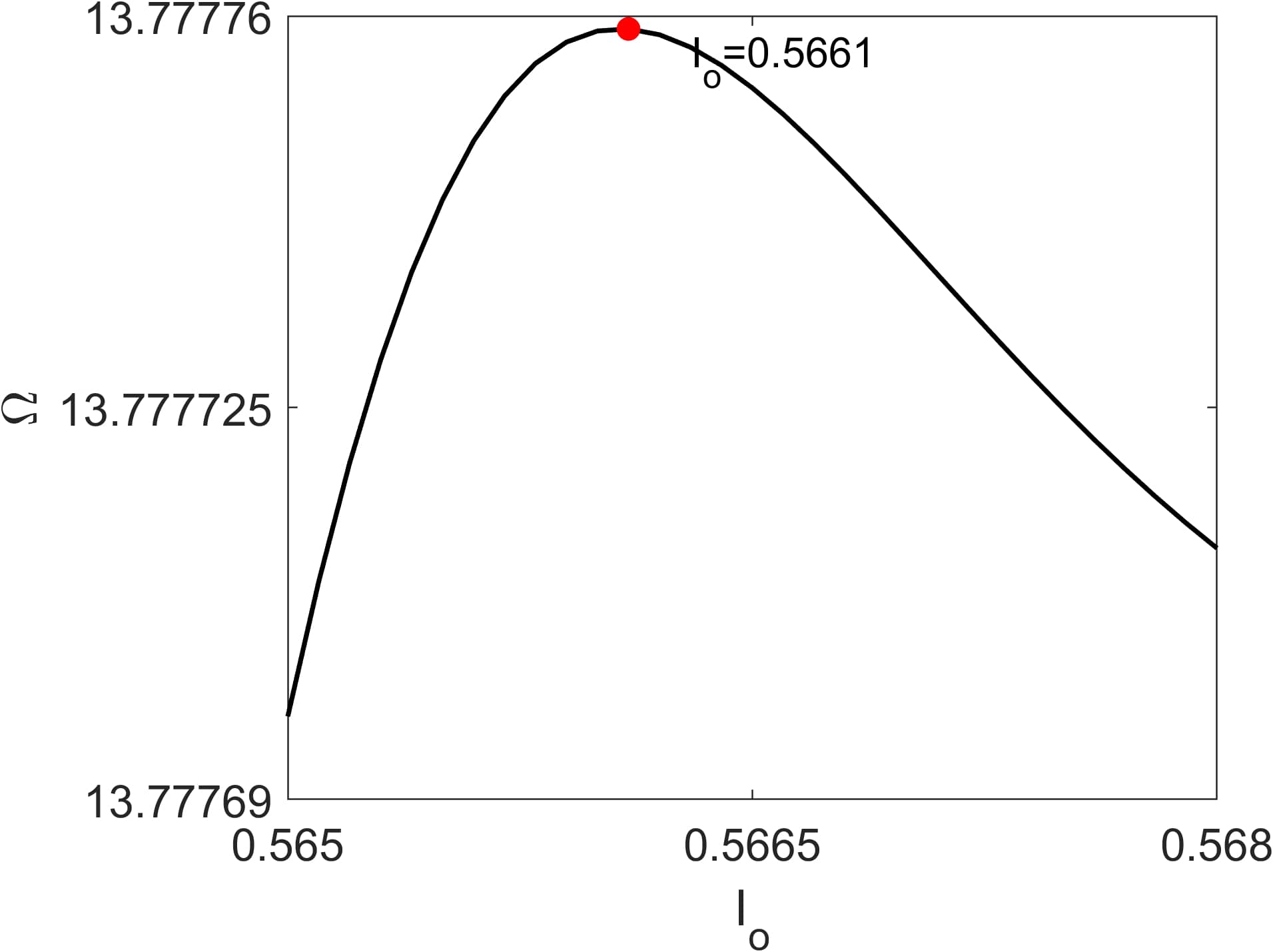}
	\caption{}
	\end{subfigure}	
    \begin{subfigure}[b]{0.35\linewidth}
	\includegraphics[width=\linewidth]{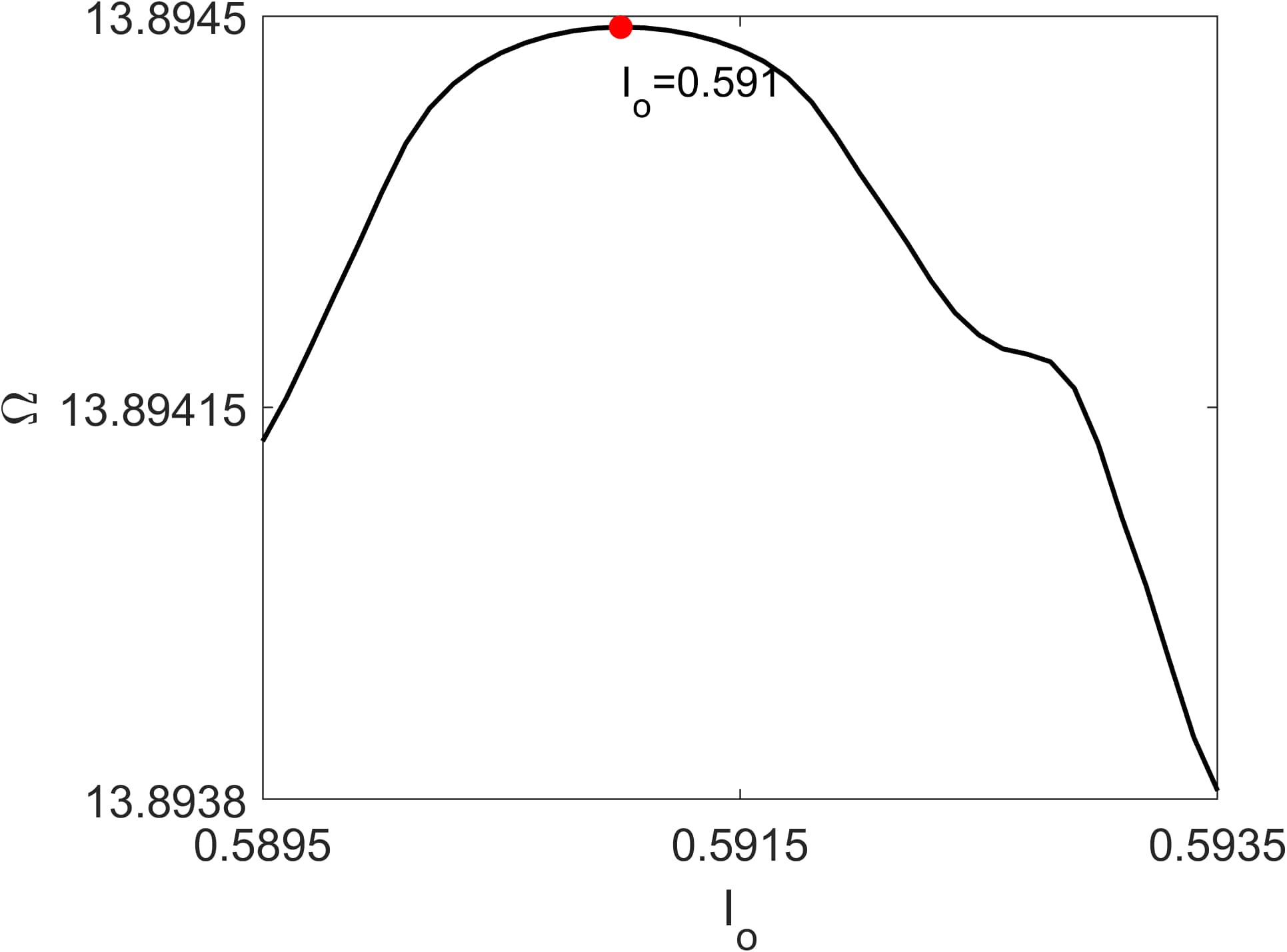}
	\caption{}
	\end{subfigure}	
  \caption{Rotation number profiles to (a) $\phi_{2}=1.5\times10^{-3}$ and (b) $1.8\times10^{-3}$ for a set of initial conditions $I{_0}$. The red dots correspond to shearless actions, indicated in the figures, required to  obtain the shearless curves of Figures \ref{fig:1}b and \ref{fig:1}d.}
  \label{fig:2}
\end{figure}

To confirm the existence of the shearless curve we present, in Figure \ref{fig:2}, the magnified rotation number profiles for Figures \ref{fig:1}b and \ref{fig:1}d, calculated for initial conditions at $\psi=0$,  with local $I$ maxima.  The $I$ maxima and the angle $\psi=0$ are used as initial coordinates and iterated to trace in red the shearless curves shown in Figures \ref{fig:1}b and \ref{fig:1}d.\\

\begin{figure}[H]
  \centering
  	\begin{subfigure}[b]{0.35\linewidth}
	\includegraphics[width=\linewidth]{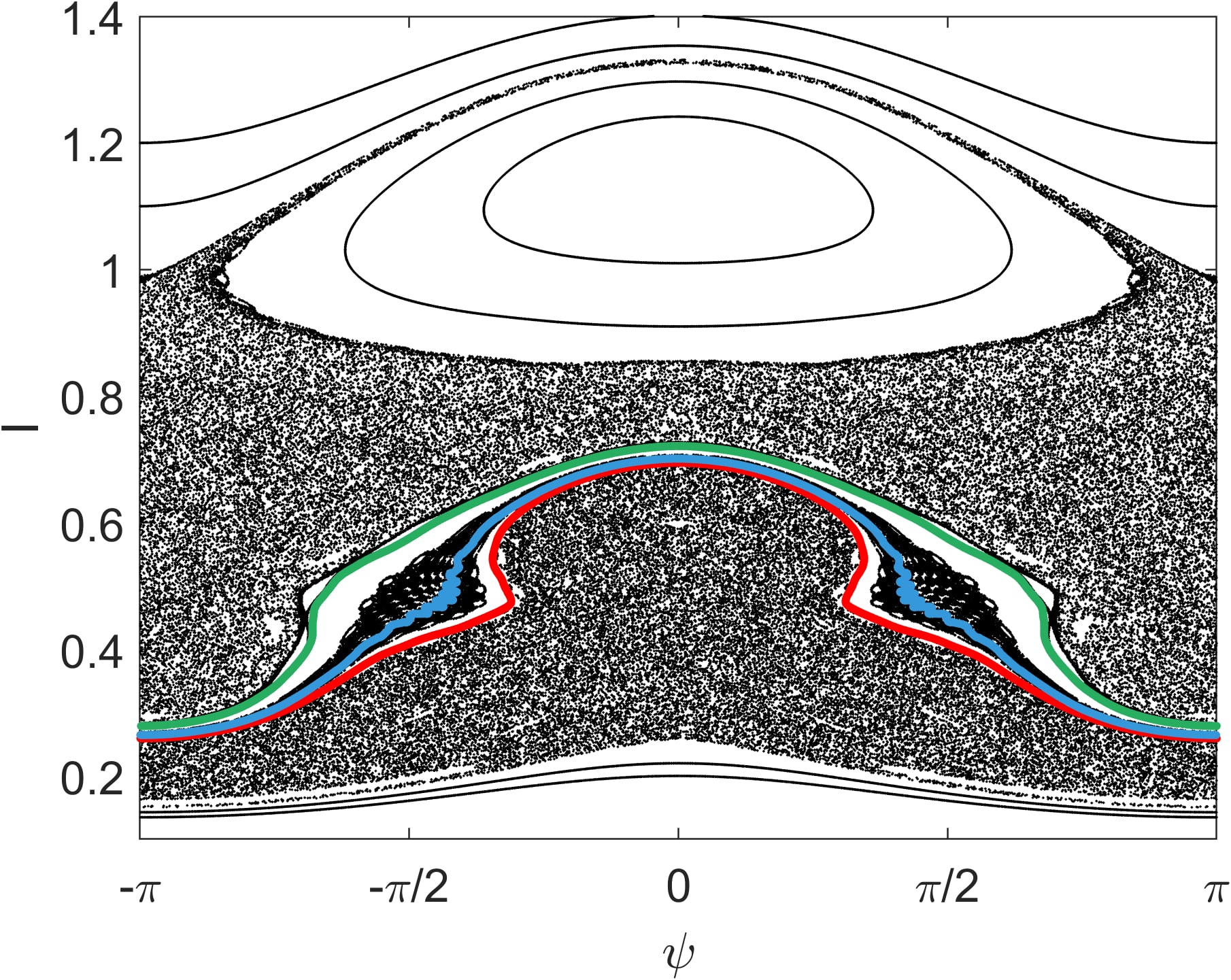}
	\caption{}
	\end{subfigure}	
    \begin{subfigure}[b]{0.35\linewidth}
	\includegraphics[width=\linewidth]{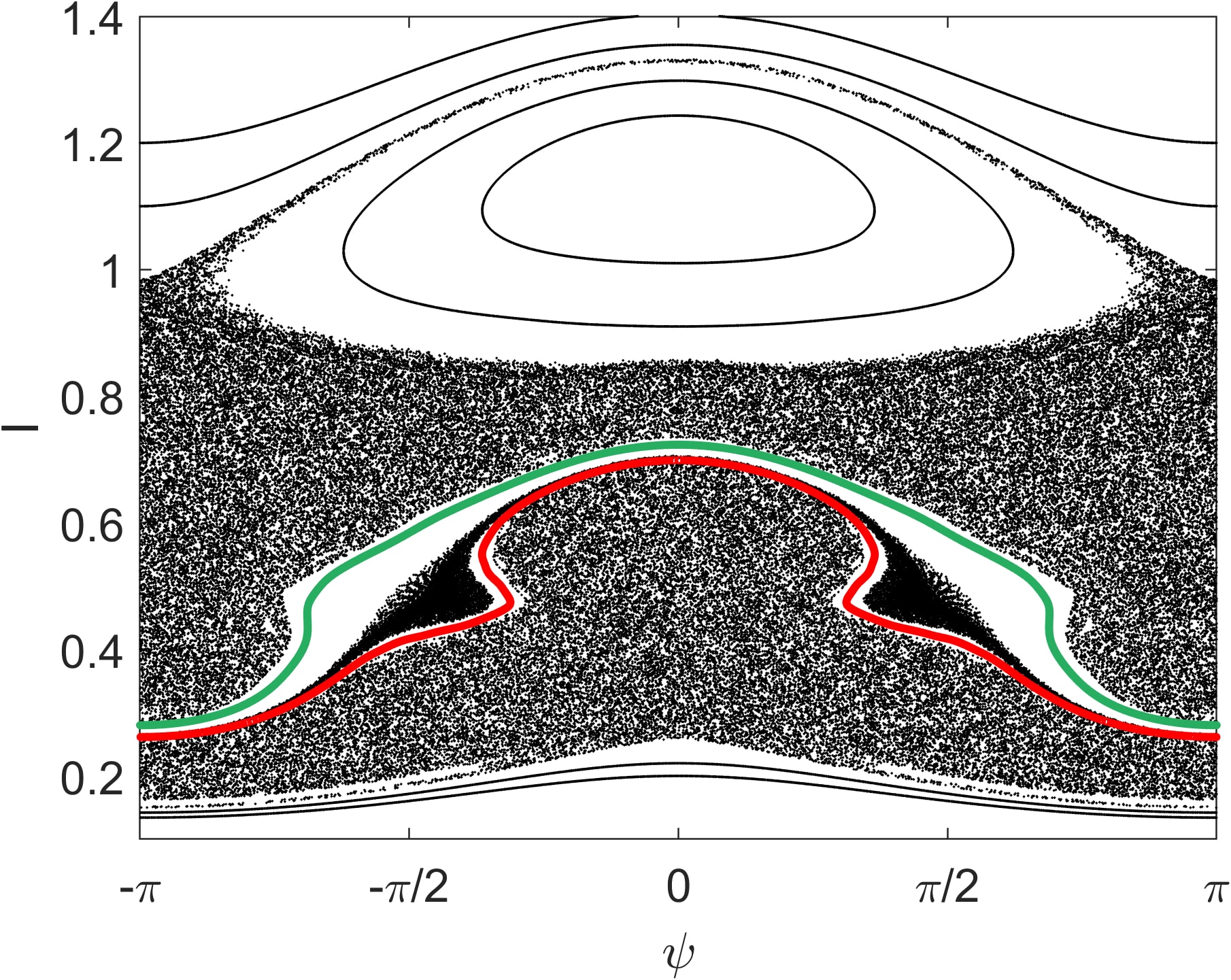}
	\caption{}
	\end{subfigure}	
	\begin{subfigure}[b]{0.35\linewidth}
	\includegraphics[width=\linewidth]{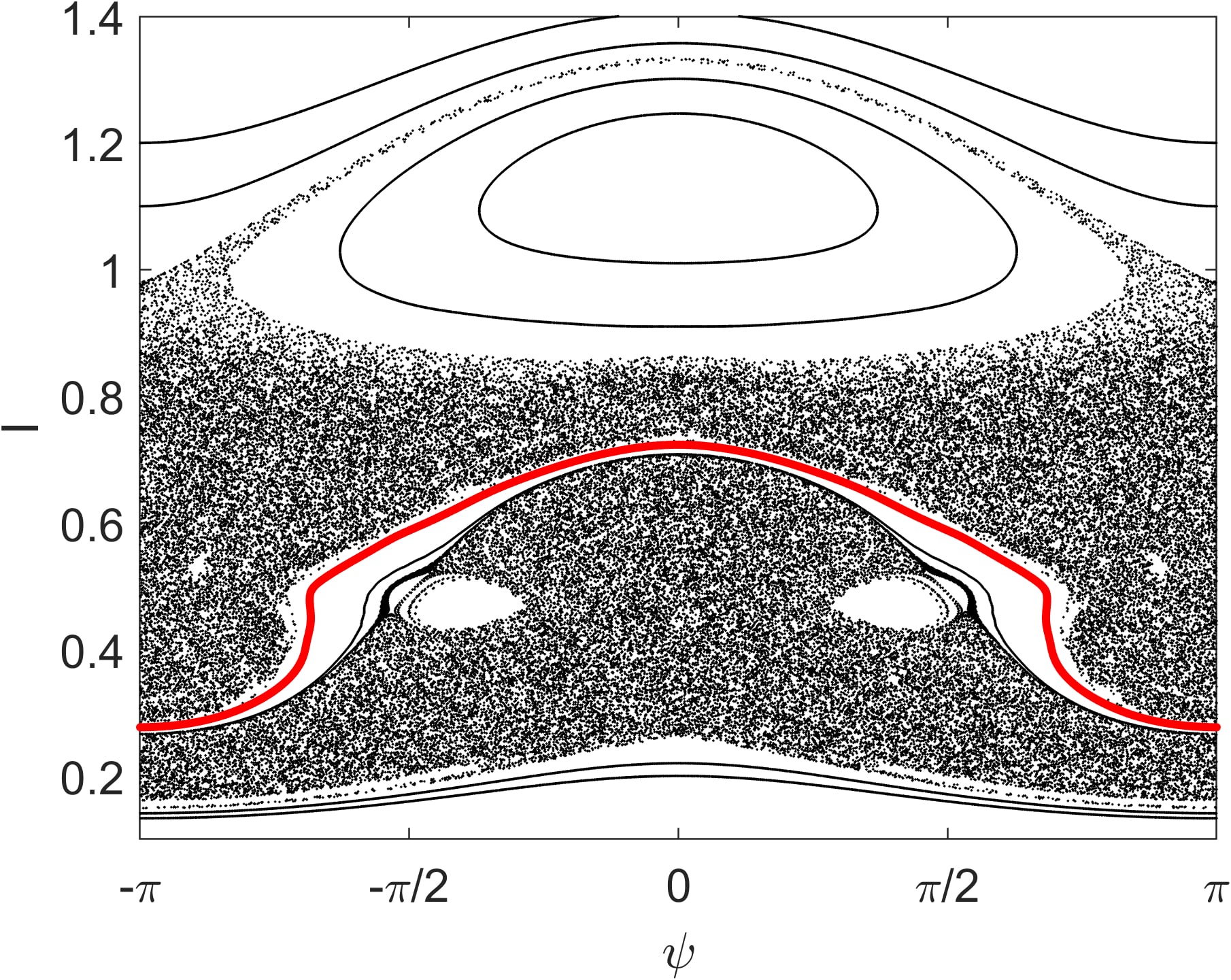}
	\caption{}
	\end{subfigure}	
    \begin{subfigure}[b]{0.35\linewidth}
	\includegraphics[width=\linewidth]{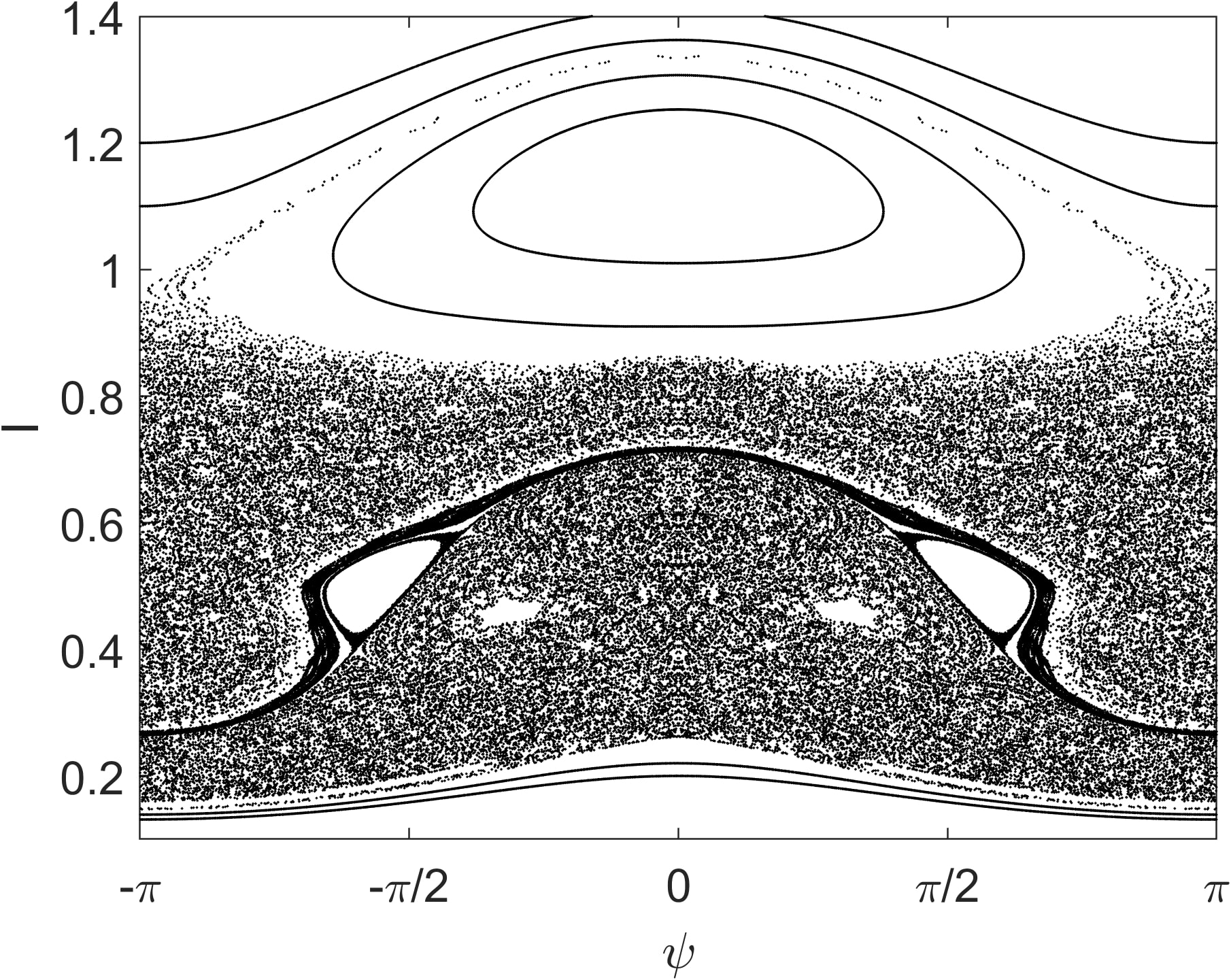}
	\caption{}
    \end{subfigure}	
  \caption{Three shearless curves, marked in red, blue and green, for (a)  $\phi_{2}=6.8\times10^{-3}$. (b) Break-up of a shearless curve, with two remaining shearless curves, marked in red and green, for $\phi_{2}=6.9\times10^{-3}$. (c) One remaining shearless curve, marked in red, for $\phi_{2}=7.1\times10^{-3}$. (d) No shearless curve, for $\phi_{2}=7.5\times10^{-3}$.}
  \label{fig:3}
\end{figure}

Next, for increasing values of $\phi_{2}$ we observe a sequence of shearless curve break-ups in phase space. Figure \ref{fig:3} shows three shearless curves for $\phi_{2}=6.8\times10^{-3}$, two shearless curves for $\phi_{2}=6.9\times10^{-3}$, one shearless curve for $\phi_{2}=7.1\times10^{-3}$ and no shearless curve for $\phi_{2}=7.5\times10^{-3}$ with a stickiness region left by the previous barrier. This kind of bifurcation has been observed in theoretical non-twist dynamical systems \cite{caroline}  and, in the context of this article, represents a reinforcement of barriers attenuating the particle transport.\\

\begin{figure}[H]
  \centering
   	\begin{subfigure}[t]{0.35\linewidth}
	\includegraphics[width=\linewidth]{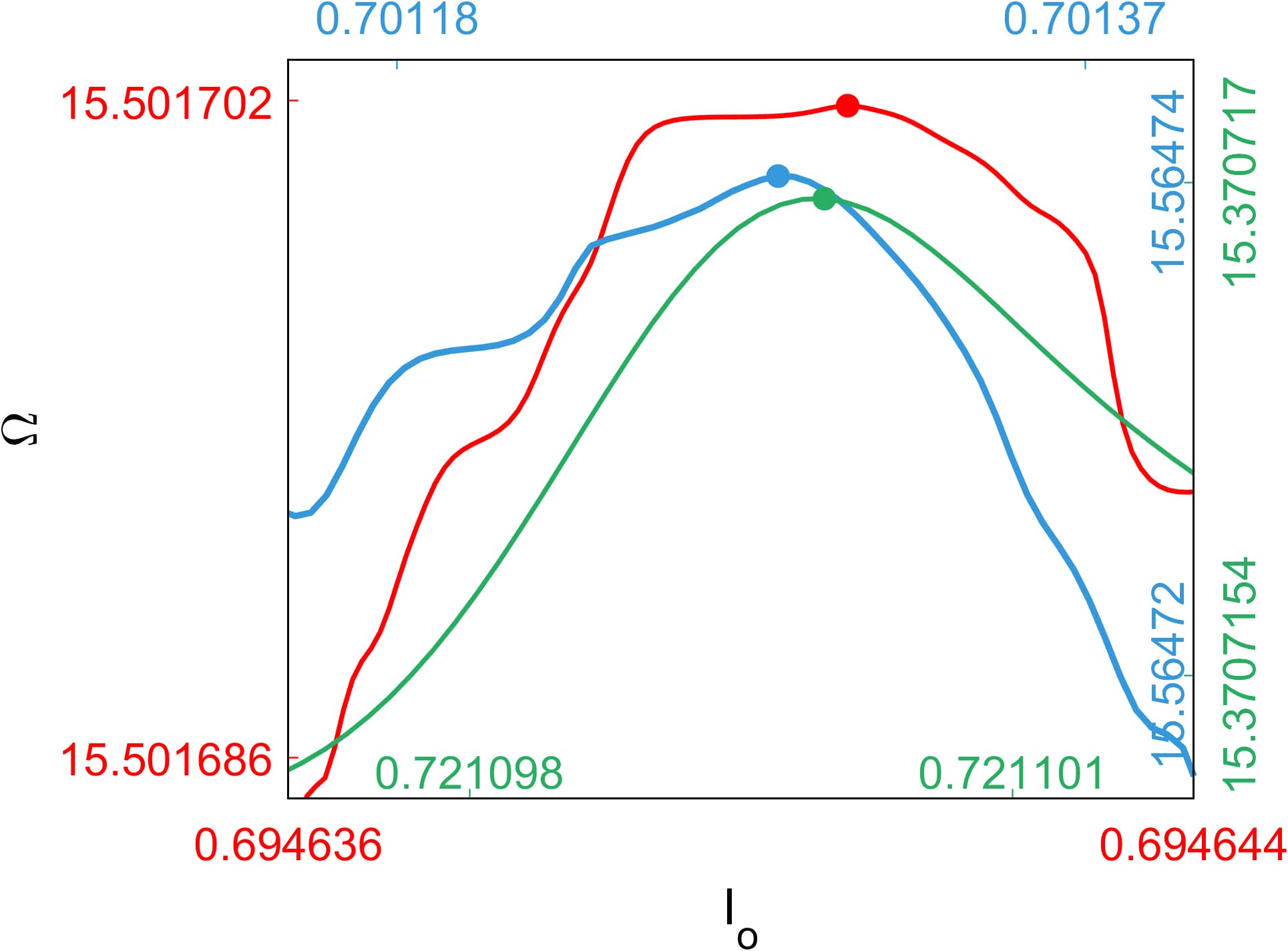}
	\caption{} 
	\end{subfigure}
    \begin{subfigure}[t]{0.35\linewidth}
	\includegraphics[width=\linewidth]{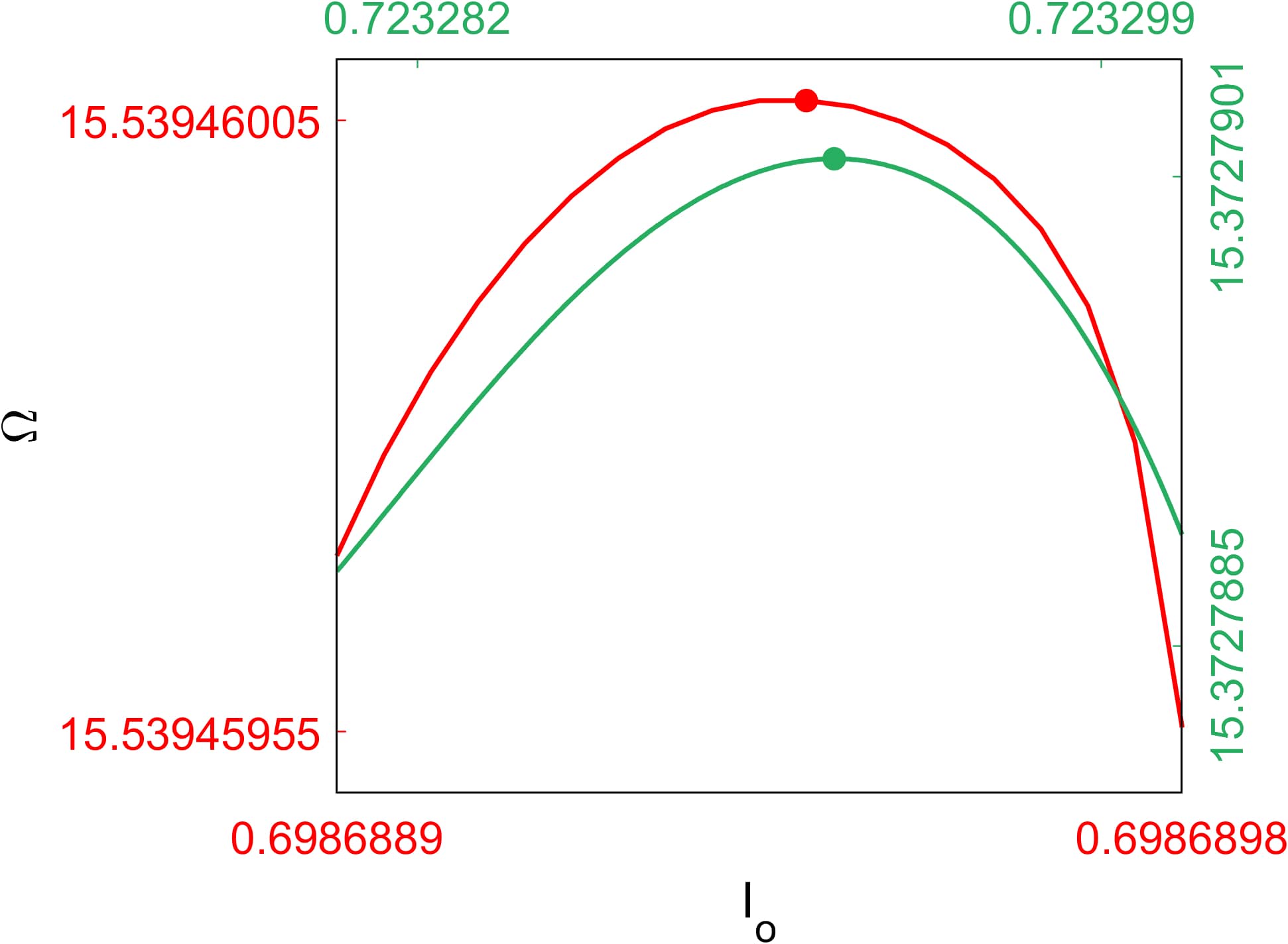}
	\caption{}
	\end{subfigure}
	\begin{subfigure}[t]{0.35\linewidth}
	\includegraphics[width=\linewidth]{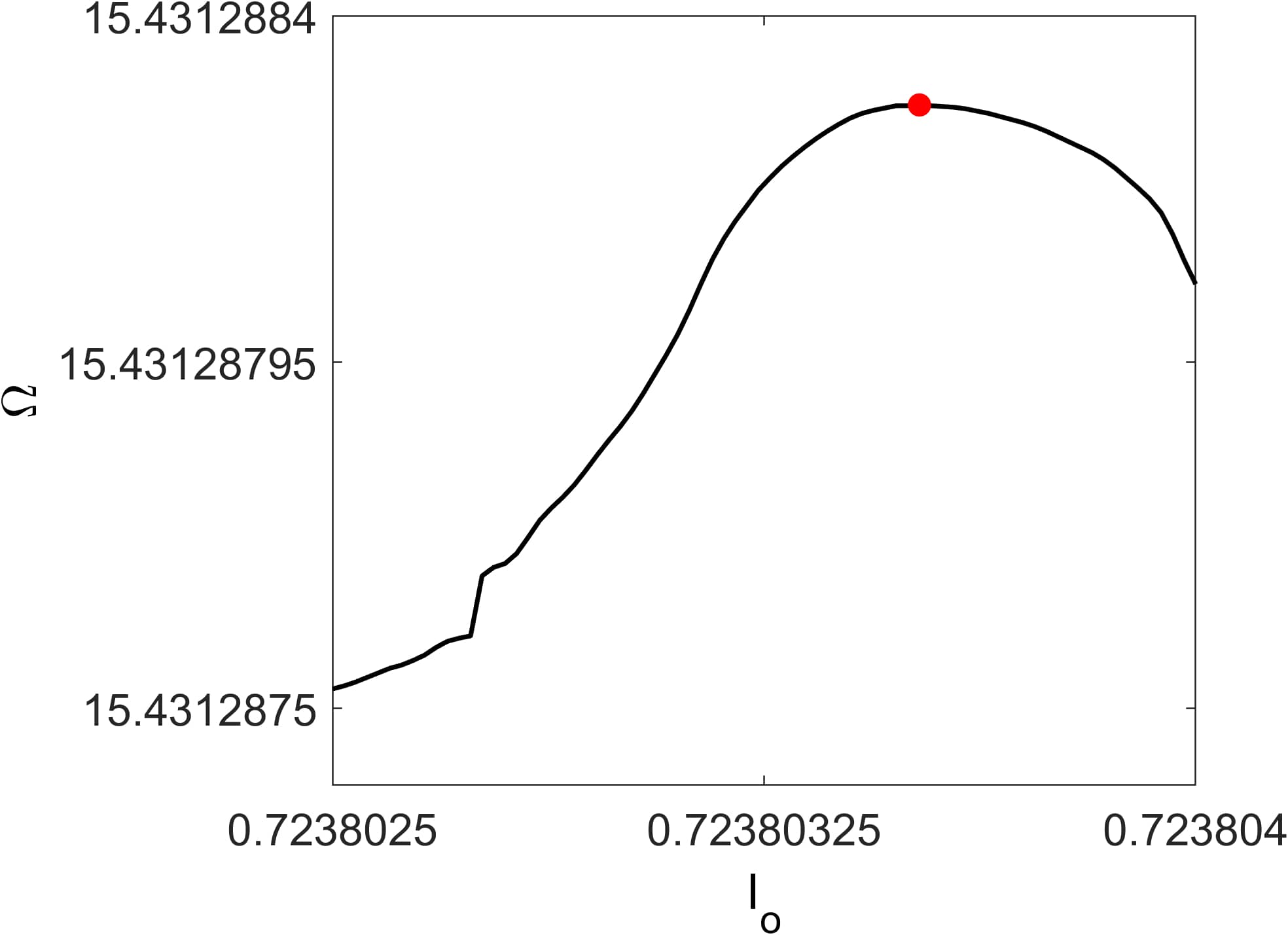}
	\caption{}
	\end{subfigure}
	\begin{minipage}[b]{0.35\linewidth}
\caption{Rotation number profiles for (a) $\phi_{2}=6.8\times10^{-3}$, (b) $\phi_{2}=6.9\times10^{-3}$ and (c) $\phi_{2}=7.1\times10^{-3}$. For each STB seen in Figure \ref{fig:3}, there is a rotation number profile in the same color as the barrier. Note that each profile, in red, blue, green and black, is associated with a unique barrier and has different $\Omega$ and $I_0$ linear scales.}\vspace{0.5cm}
\label{fig:4}
\end{minipage}
\end{figure}

For each coloured STB seen in Figure \ref{fig:3}, we present, in Figure \ref{fig:4}, the associated magnified rotation number profiles in the same color as the STB curves, calculated in the non chaotic map region, for initial conditions with $\psi = 0$. These STBs can be generated by the initial coordinates $\psi = 0$ and $I$ corresponding to the local maxima indicated in the profiles of Figure \ref{fig:4}.\\

\begin{figure}[H]
  \centering
  \begin{subfigure}[b]{0.35\linewidth}
	\includegraphics[width=\linewidth]{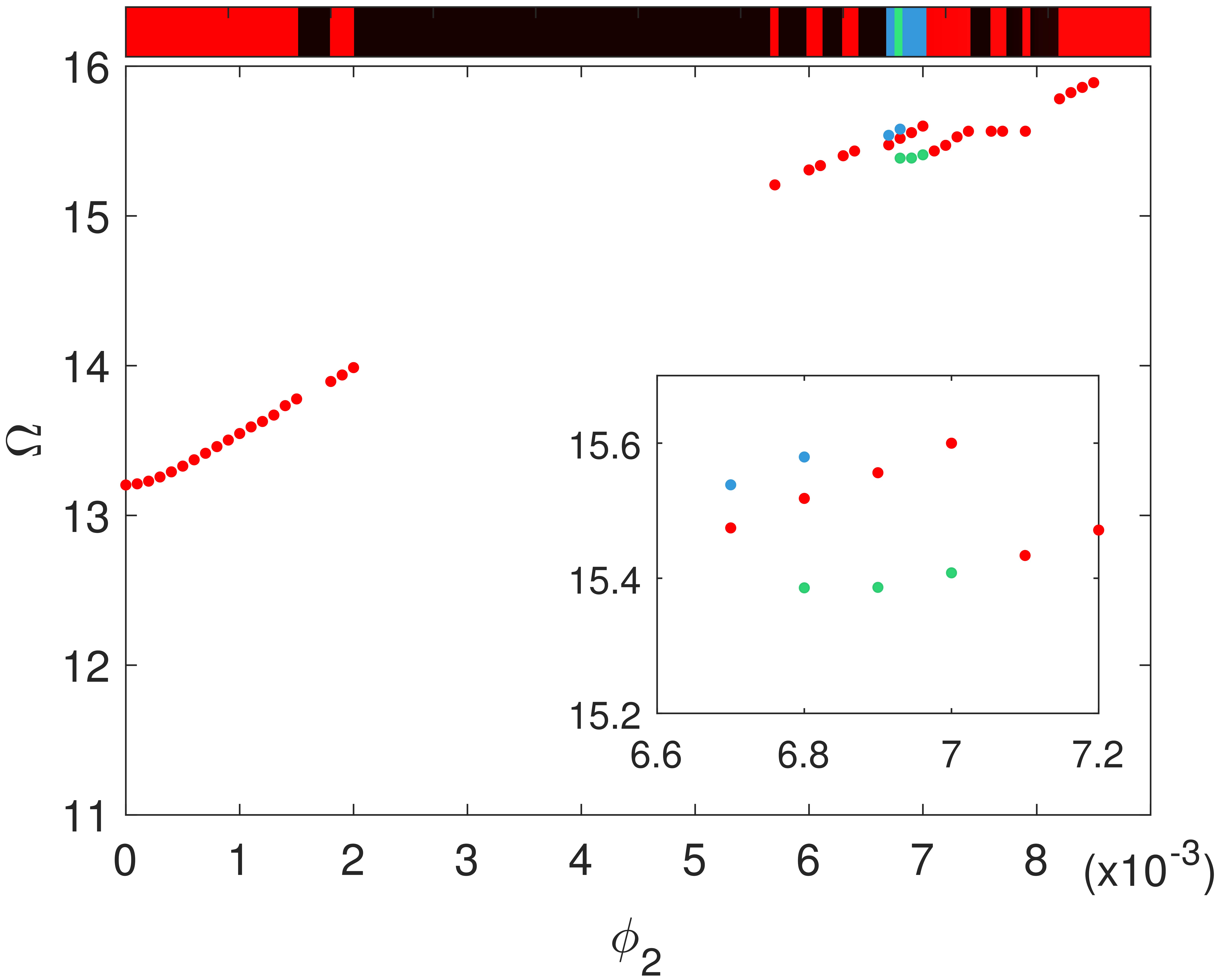}
	\caption{}
	\end{subfigure}	
    \begin{subfigure}[b]{0.36\linewidth}
	\includegraphics[width=\linewidth]{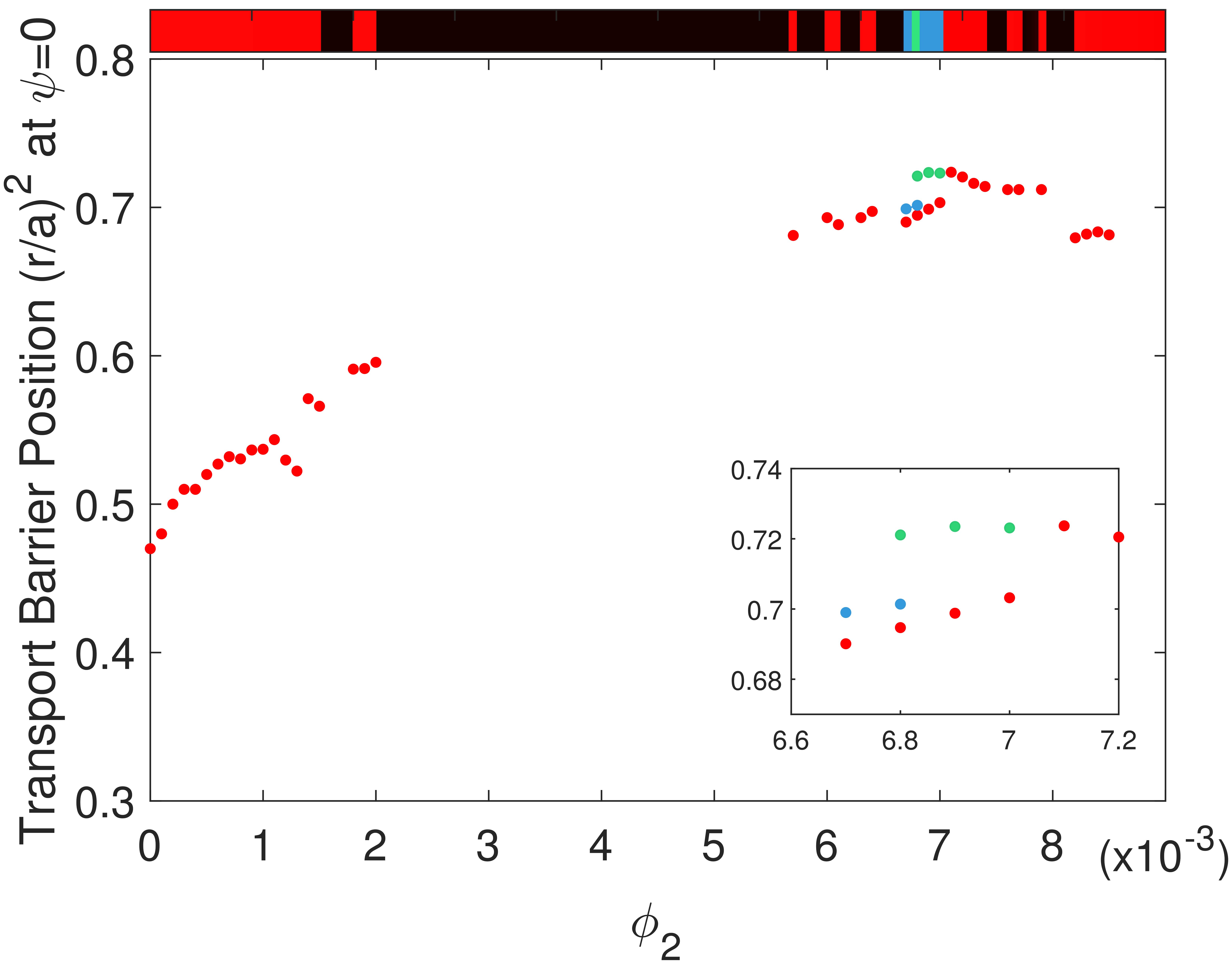}
	\caption{}
	\end{subfigure}	
  \caption{(a) Rotation number $\Omega$ as function of the fluctuation amplitude $\phi_{2}$. Red (black) bars indicate the existence (non existence) of shearless barriers. For some values of  $\phi_{2}$, there are two or three shearless curves, indicated in blue or green bars.
(b) Transport barrier position $I$ at $\psi=0$ versus $\phi_{2}$.}
  \label{fig:5}
\end{figure}

To show that the existence of barriers depends on the fluctuation amplitude $\phi_{2}$, we check their existence in an interval of $\phi_{2}$ from $0$ to $8.5\times10^{-3}$, with a step $0.1\times10^{-3}$. The results are shown in Figure \ref{fig:5}. In this figure, the red, blue and green bars represent the amplitude range for which barriers exist, otherwise we draw a black bar, when the barrier is not observed. We also present magnifications to better show the shearless bifurcations changing the number of barriers. Figure \ref{fig:5}a shows the rotation number as a function of fluctuation amplitude and Figure \ref{fig:5}b shows the shearless barrier position as a function of $\phi_{2}$.  We conclude that, as the amplitude of the fluctuation varies, the shearless barriers can be recurrently destroyed and restored. It means that the existence of barriers, in tokamak discharges, may depend on the fluctuation amplitude. In addition, for some fluctuation ranges, there are barrier bifurcations giving rise to a second and, possibly, a third barrier. 

\section{Persistent Barriers\label{sec:Persistent-Barriers}}

\begin{figure}[H]
  \centering
  	\begin{subfigure}[b]{0.35\linewidth}
	\includegraphics[width=\linewidth]{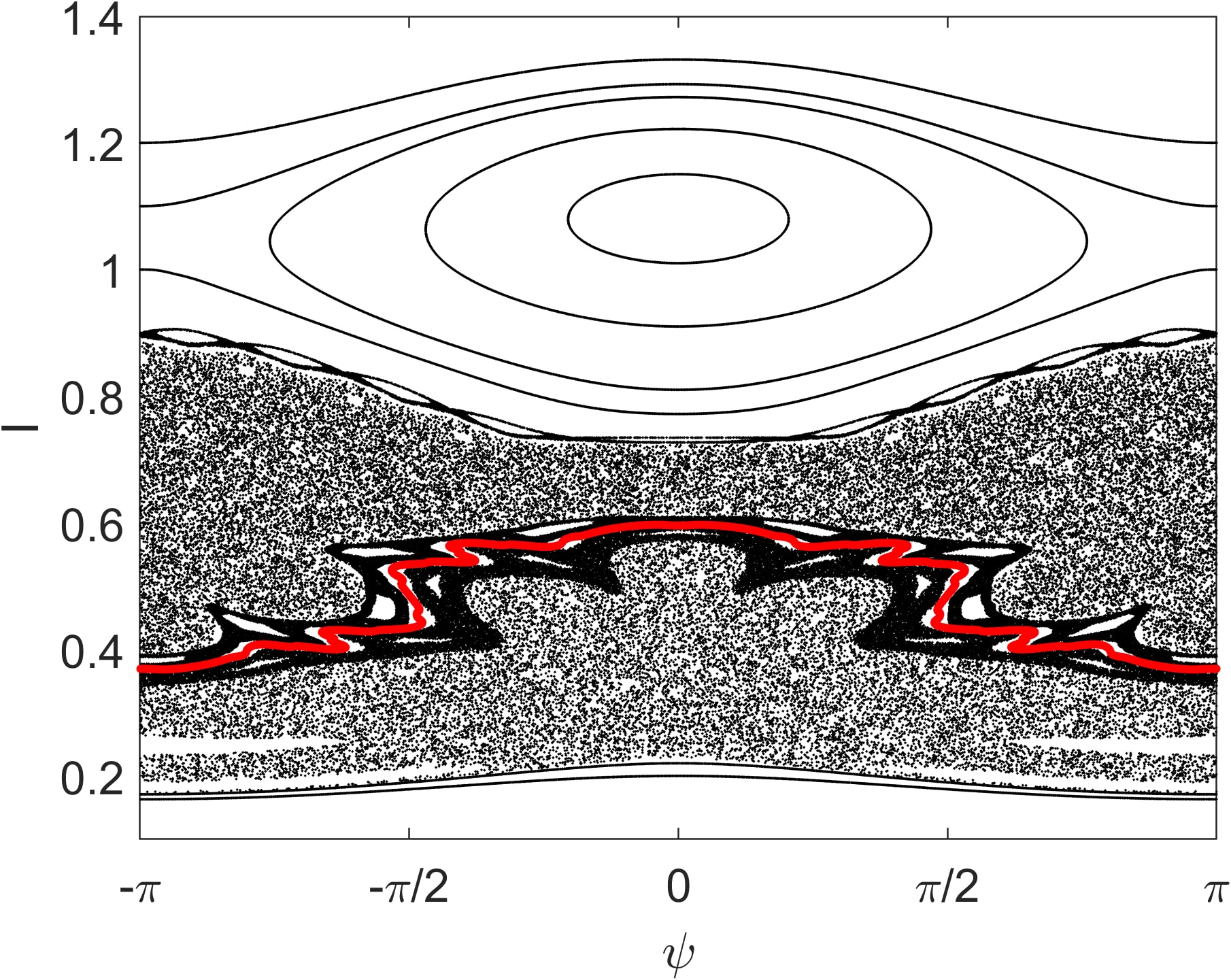}
	\caption{}
	\end{subfigure}	
    \begin{subfigure}[b]{0.35\linewidth}
	\includegraphics[width=\linewidth]{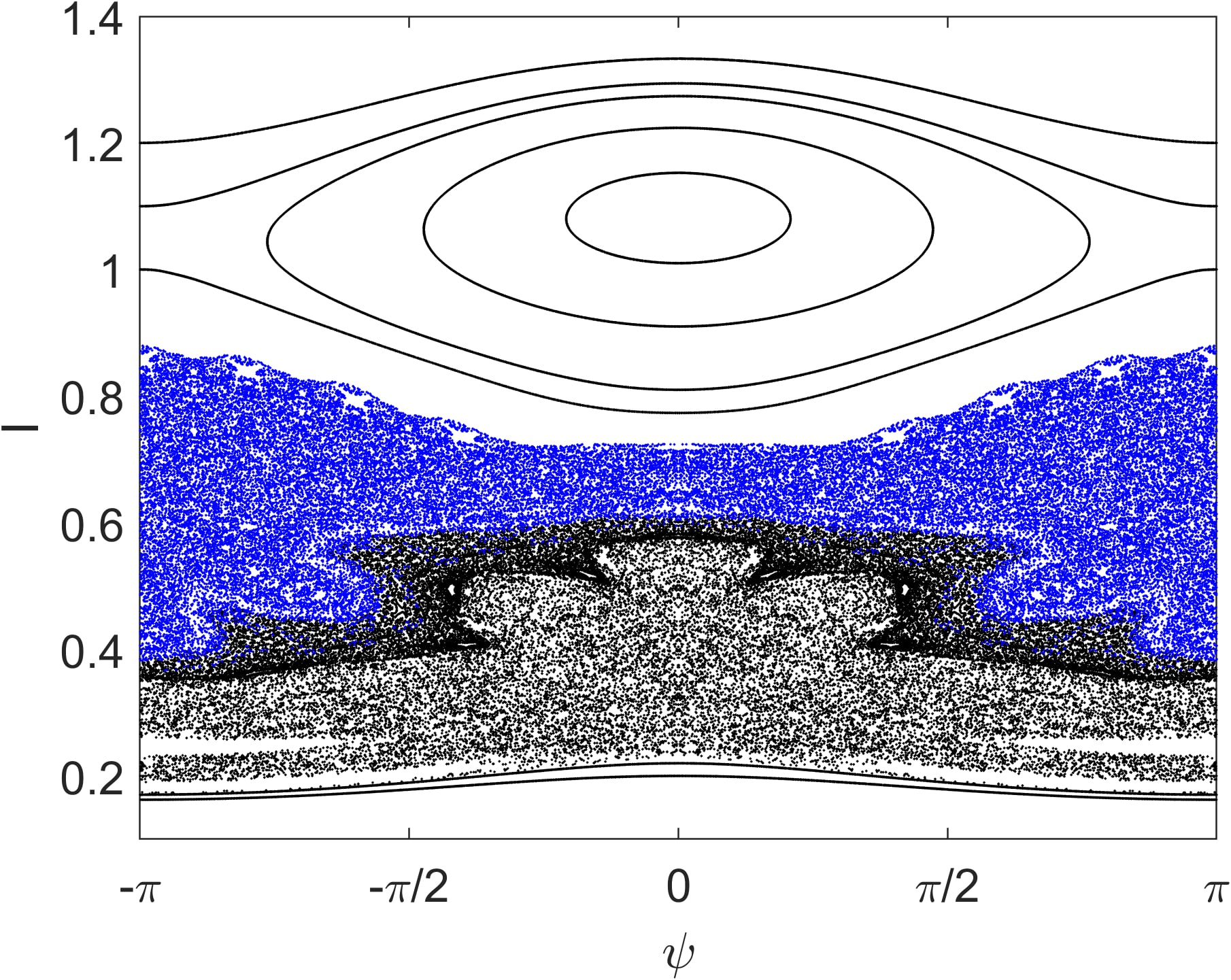}
	\caption{}
	\end{subfigure}	
  \caption{(a) Shearless curve, in red color, for $\phi_{2}=2.0\times10^{-3}$  and the stickiness appearance around this curve. (b) For $\phi_{2}=2.1\times10^{-3}$, the shearless curve is destroyed and stickiness region expanded. Two chaotic orbits in blue and black are concentrated in the stickiness region and take a long time to cross this region. 
  The stickiness regions are identified by the high concentration of black points.} 
  \label{fig:6}
\end{figure}

A noticeable feature in non-twist systems is the appearance of a layer of stickiness after the shearless barrier break-up or before its onset \cite{santos}. A stickiness layer separates the chaotic region in two parts and, to cross this layer, orbits get trapped for a high number of iterations \cite{szezech2}. Accordingly, for some domains of the amplitude perturbation, in the present analysis we observe that stickiness regions appear in phase space, associated to the shearless barrier break-up or to its onset. This is the case for  $\phi_{2}=2.0\times10^{-3}$. Taking $\phi_{2}=2.1\times10^{-3}$, the barrier is destroyed and a stickiness region is magnified, as shown in Figure \ref{fig:6}, separating the chaotic orbits around the stickiness layer.\\

In Figure \ref{fig:6}b, one chaotic orbit in blue and another one in black fill separated areas in phase space, with a high concentration of iterations in the stickiness region, spending a long time to cross this region. The possible crossings are too long to be observed in Figure \ref{fig:6}b. Thus, even if the barrier is broken up, there is a long stickiness inhibiting  the particle transport in this region indicating the persistence of barrier effect, even after the barrier disappearance.

\begin{figure}[H]
  \centering
  	\begin{subfigure}[b]{0.35\linewidth}
	\includegraphics[width=\linewidth]{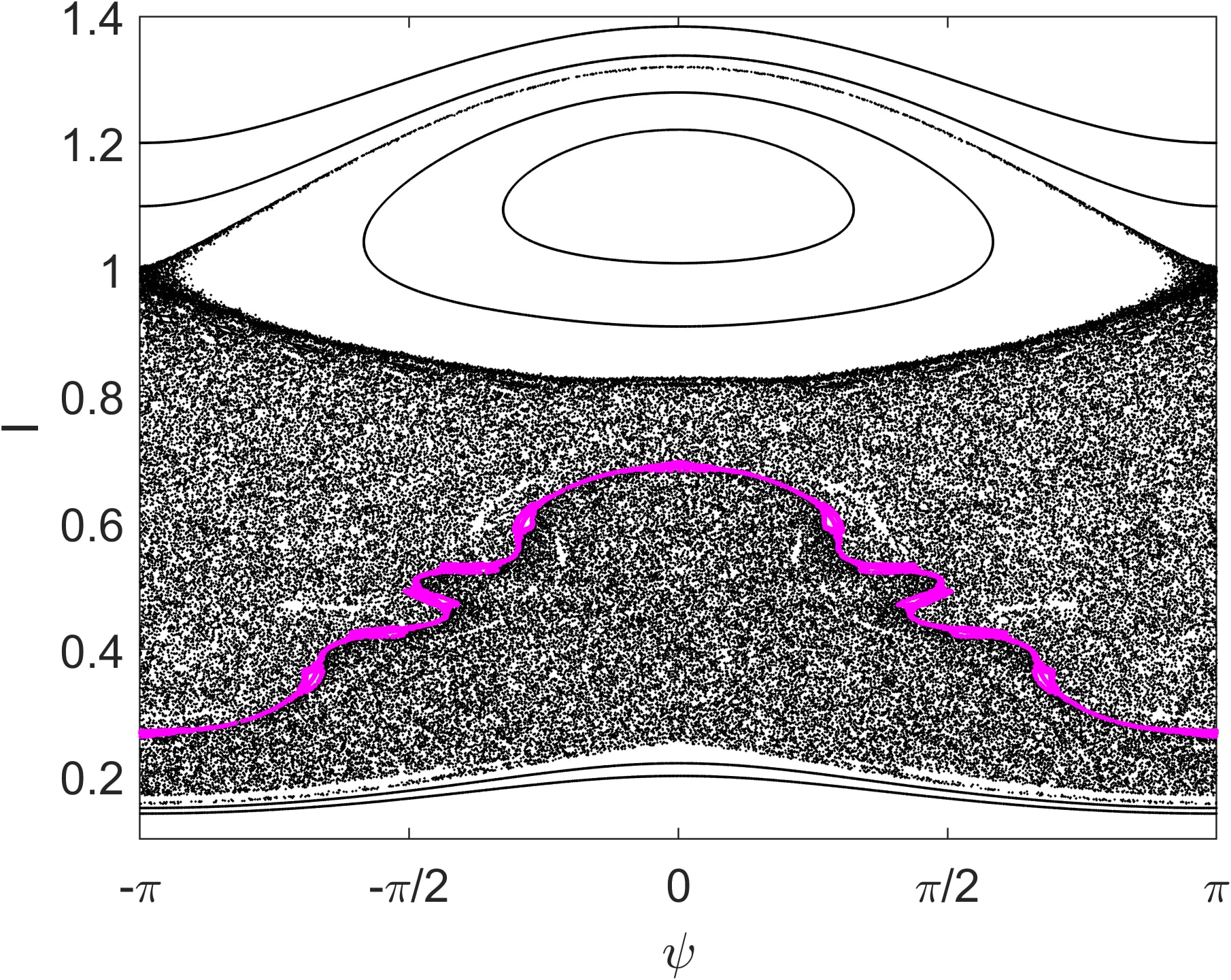}
	\caption{}
	\end{subfigure}	
    \begin{subfigure}[b]{0.35\linewidth}
	\includegraphics[width=\linewidth]{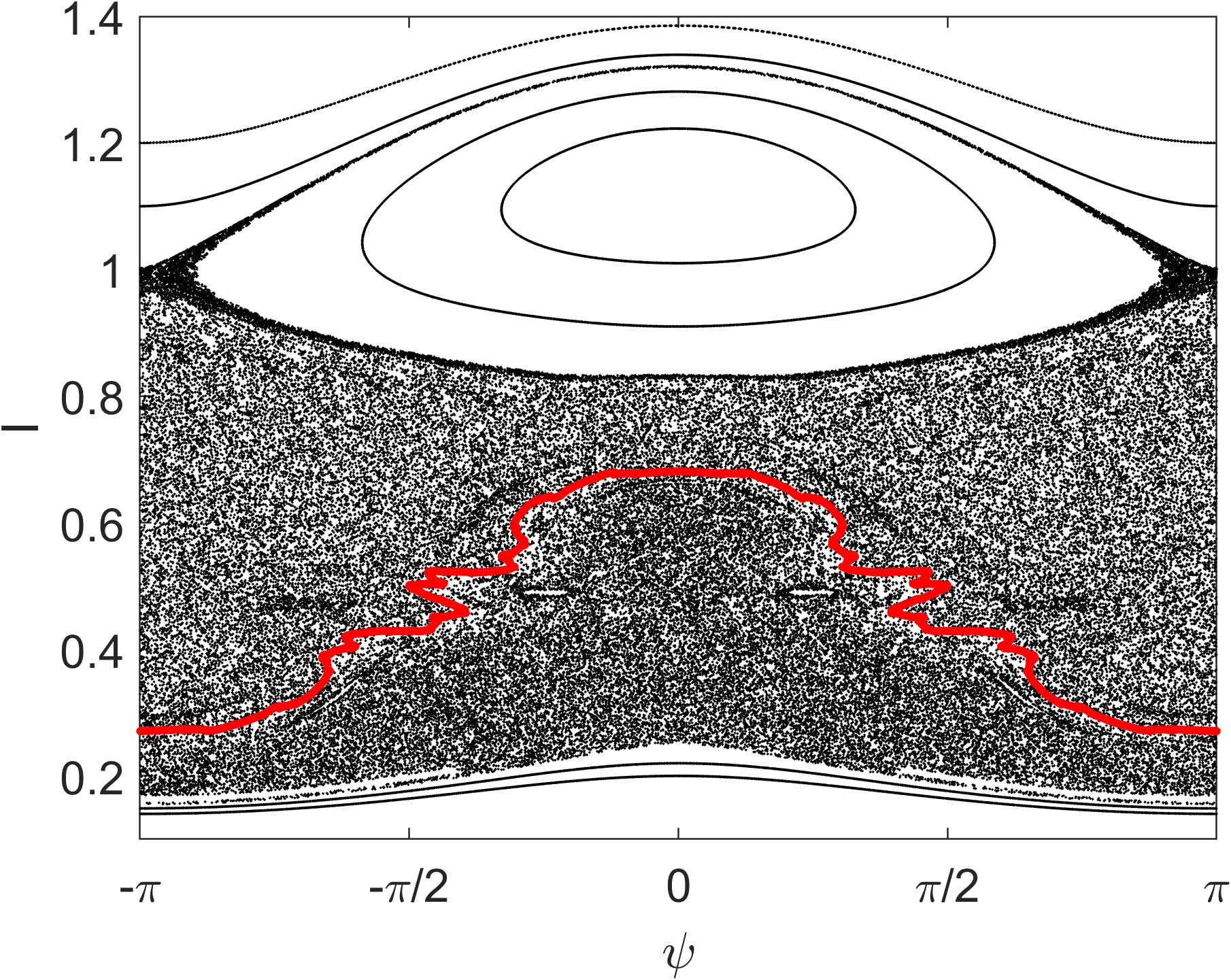}
	\caption{}
	\end{subfigure}	
   \caption{(a) Emergence of stickiness, precursor of the shearless curve for  $\phi_{2}=5.6\times10^{-3}$. The stickiness layer is represented in magenta, where the shearless curve is created.
(b) Shearless curve onset in the stickiness region, for       $\phi_{2}=5.7\times10^{-3}$, in red.}
  \label{fig:7}
\end{figure}

Figure \ref{fig:7} shows the emergence of stickiness, precursor of the shearless curve, for  $\phi_{2}=5.6\times10^{-3}$. 
For $\phi_{2}=5.7\times10^{-3}$, the shearless curve onset occurs in the stickiness layer. This transformation is the reverse sequence of that shown in Figure \ref{fig:6}, namely, the stickiness appears before the barrier onset and not after its destruction.  It also indicates the persistence of the barrier effect after this disappearance for decreasing fluctuation amplitude.

\begin{figure}[H]
  \centering
  \begin{subfigure}[b]{0.35\linewidth}
	\includegraphics[width=\linewidth]{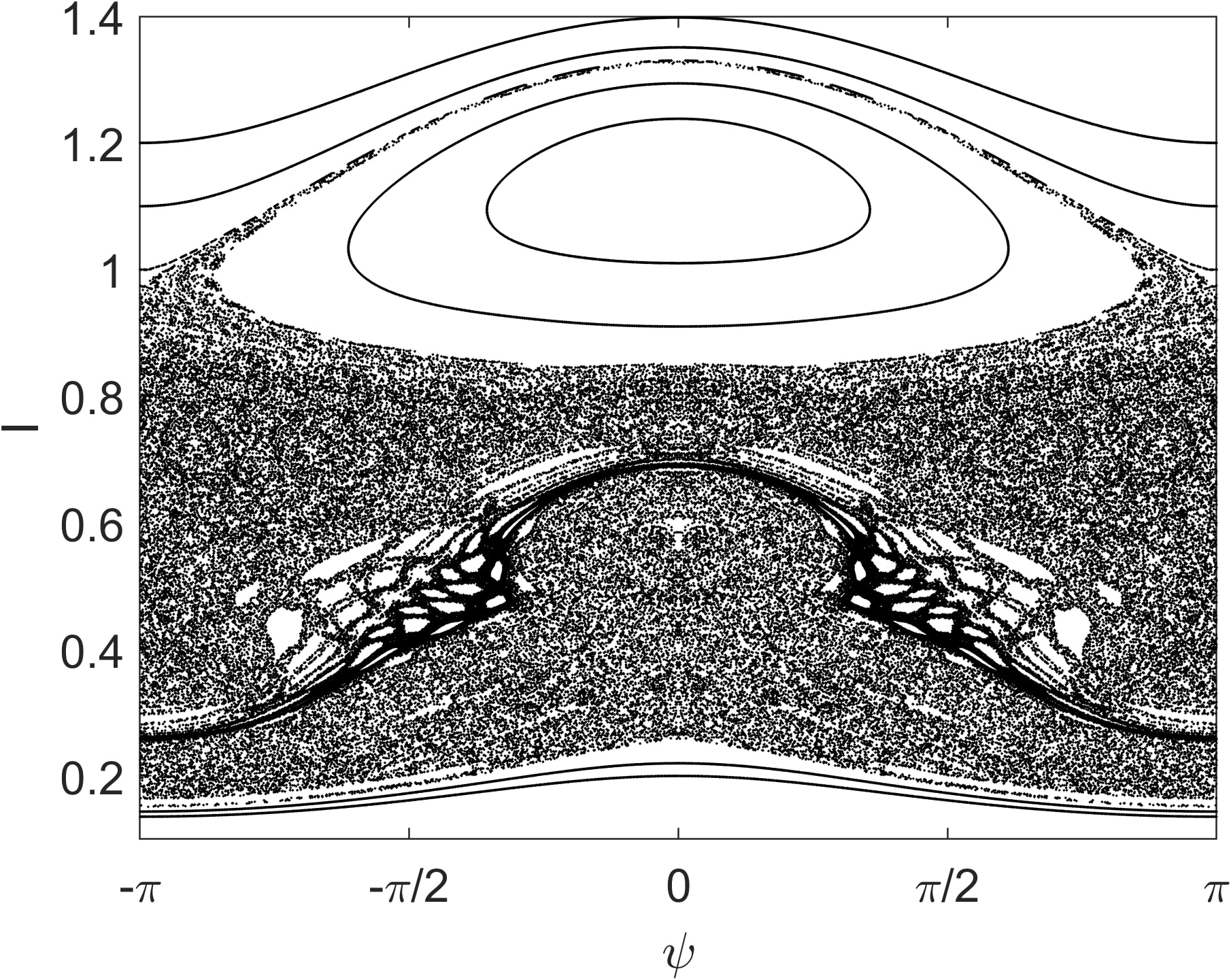}
	\caption{}
	\end{subfigure}	
	\begin{subfigure}[b]{0.35\linewidth}
	\includegraphics[width=\linewidth]{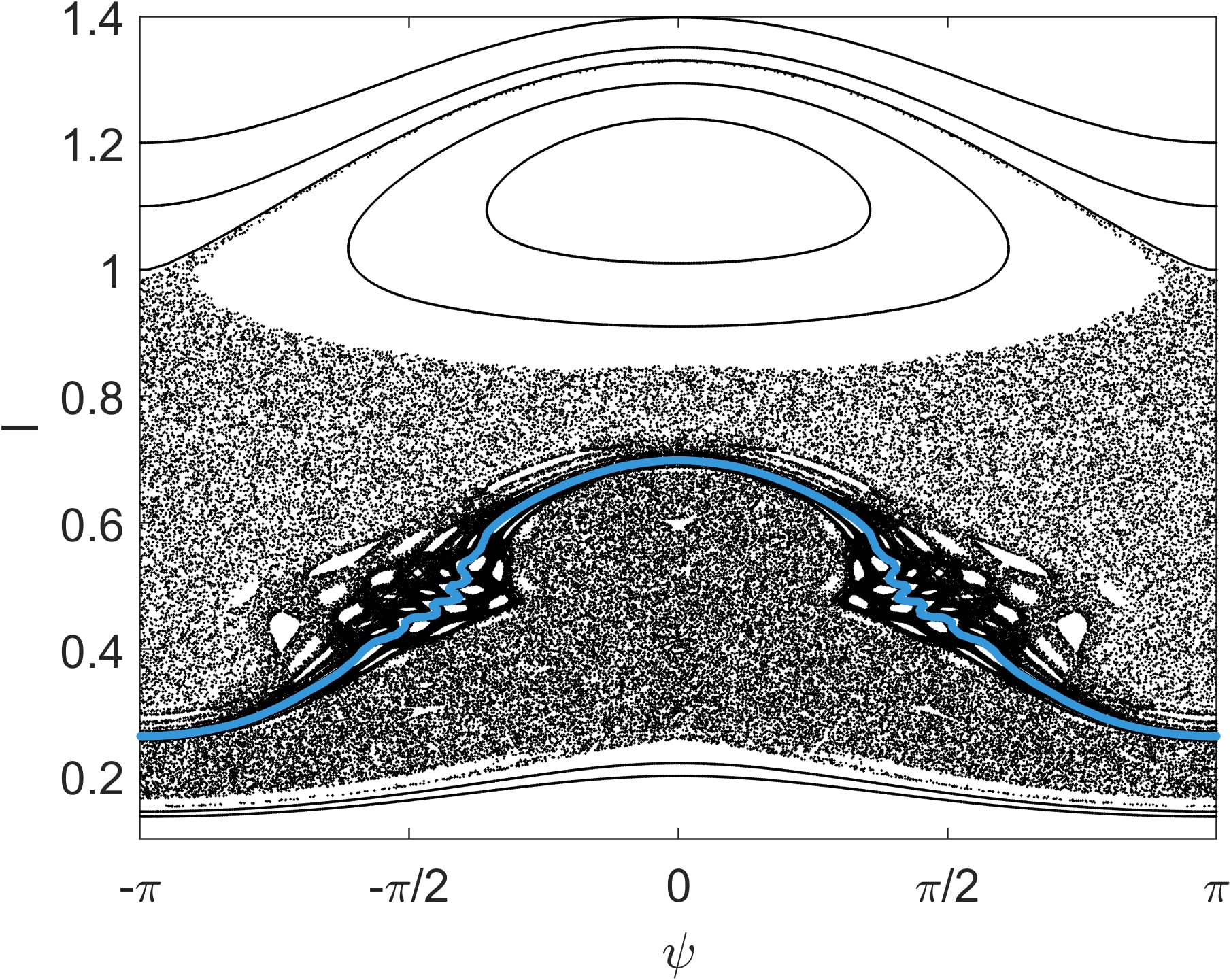}
	\caption{}
	\end{subfigure}	
    \begin{subfigure}[b]{0.35\linewidth}
	\includegraphics[width=\linewidth]{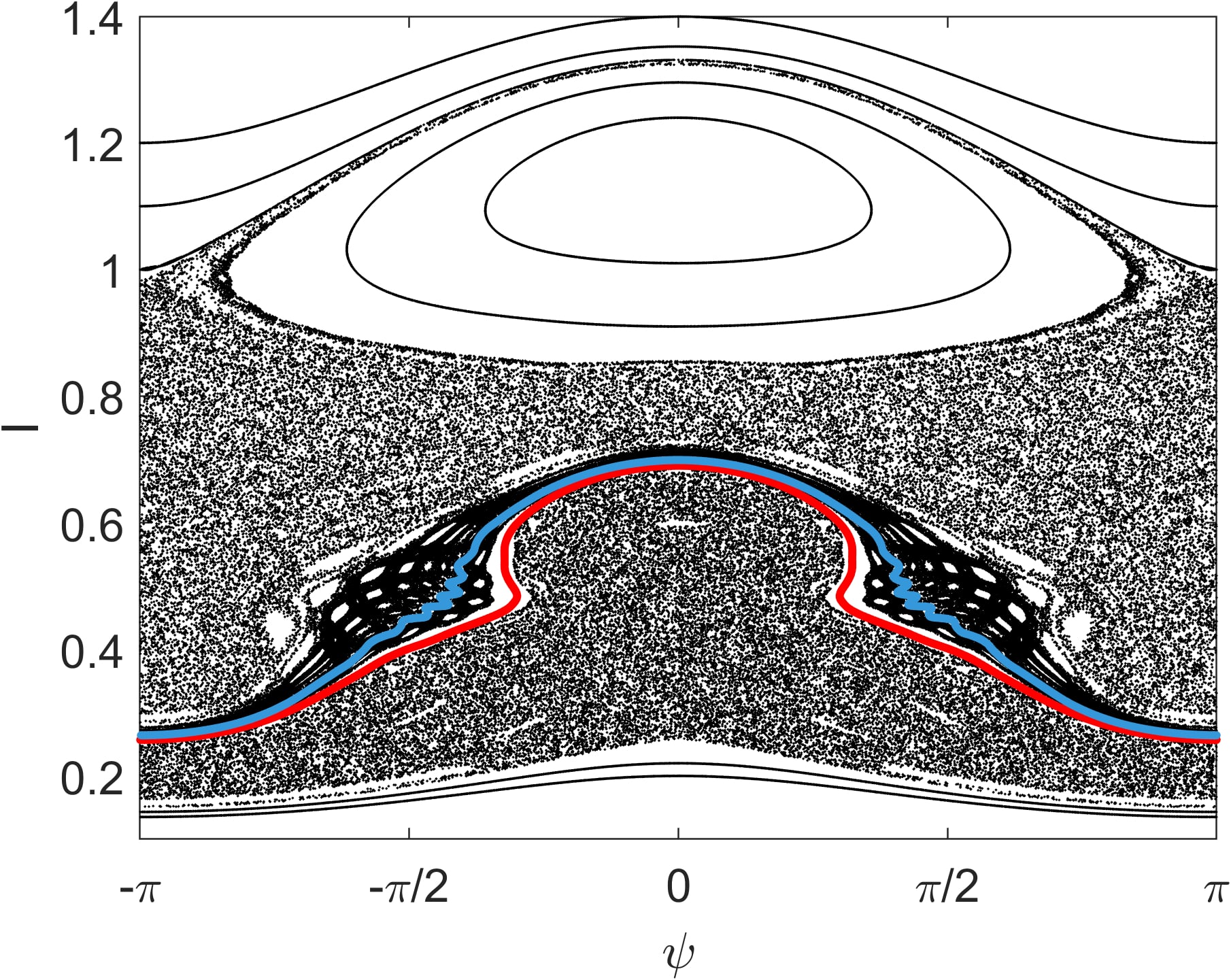}
	\caption{}
	\end{subfigure}	
  \caption{(a) No shearless curve for $\phi_{2}=6.6\times10^{-3}$.  (b) Emerging shearless curve for $\phi_{2}=6.61\times10^{-3}$, in blue color. (c) Two shearless barriers for $\phi_{2}=6.7\times10^{-3}$, in red and blue.}    
  \label{fig:8}
\end{figure}

As we show in Figure \ref{fig:8}, increasing the perturbation amplitude, for $\phi_{2}=6.6\times10^{-3}$, there is no shearless curve. However, for $\phi_{2}=6.61\times10^{-3}$ a shearless curve appears and then, for $\phi_{2}=6.7\times10^{-3}$, we see two shearless curves (in blue and red). In this Figure, we have an example of a large region of stickiness, with many visible islands, associated to the onset of multiple shearless barriers.

\section{Influence of the radial electric field profile on the shearless transport barriers\label{sec:fixed phi2}}

The experiments show that the transport barriers appear by modifications of the radial electric field $E_r$ in high confinement regime, as discussed in the Introduction. On the other hand, shearless barriers appear as
a mechanism to prevent chaotic particle transport: even after the invariant curve has been broken, the islands still remain with large stickiness that reduce the transport.\\
 
\begin{figure}[H]
  \centering
   \begin{subfigure}[b]{0.33\linewidth}
	\includegraphics[width=\linewidth]{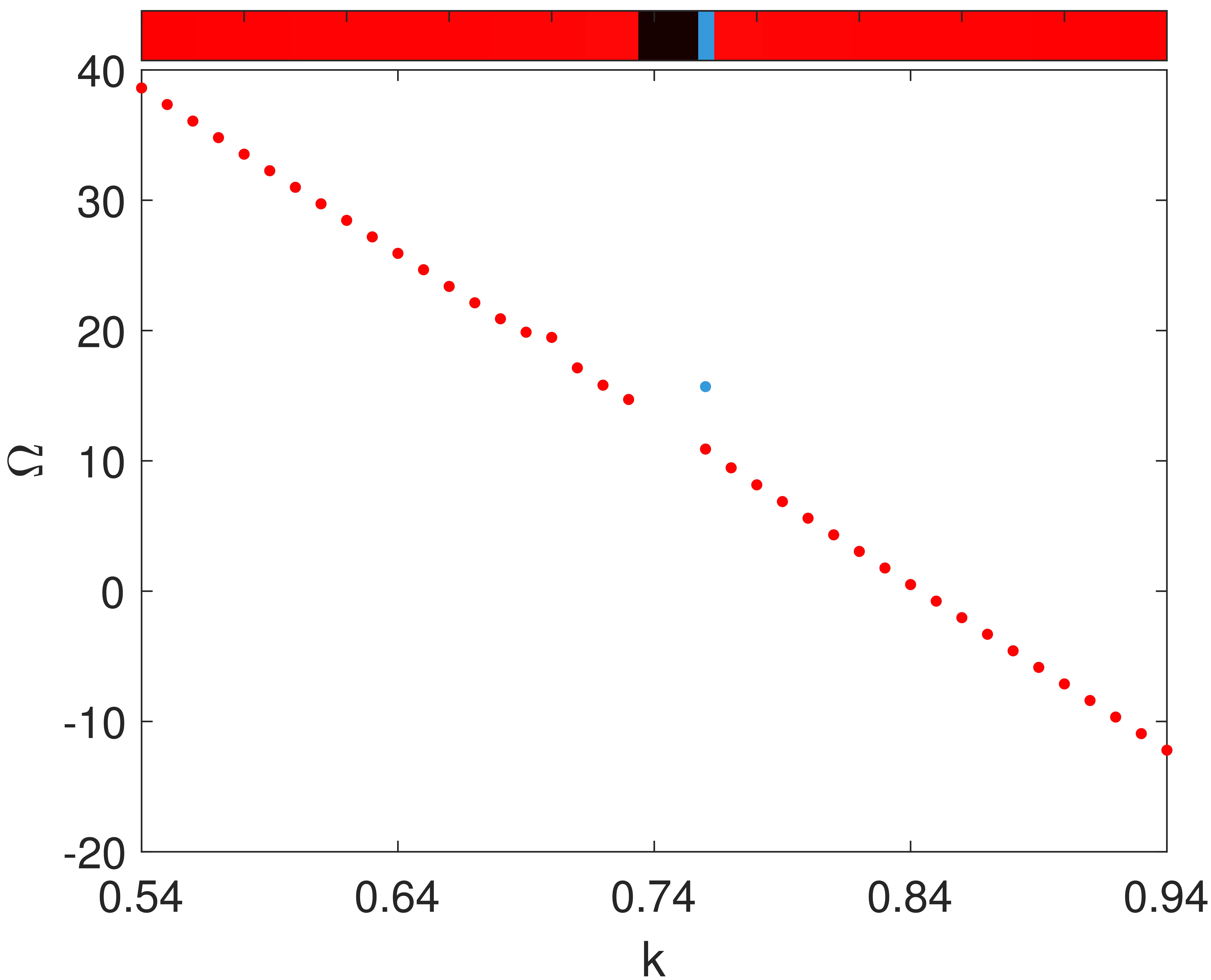}
	\caption{}
	\end{subfigure}	
	\begin{subfigure}[b]{0.33\linewidth}
	\includegraphics[width=\linewidth]{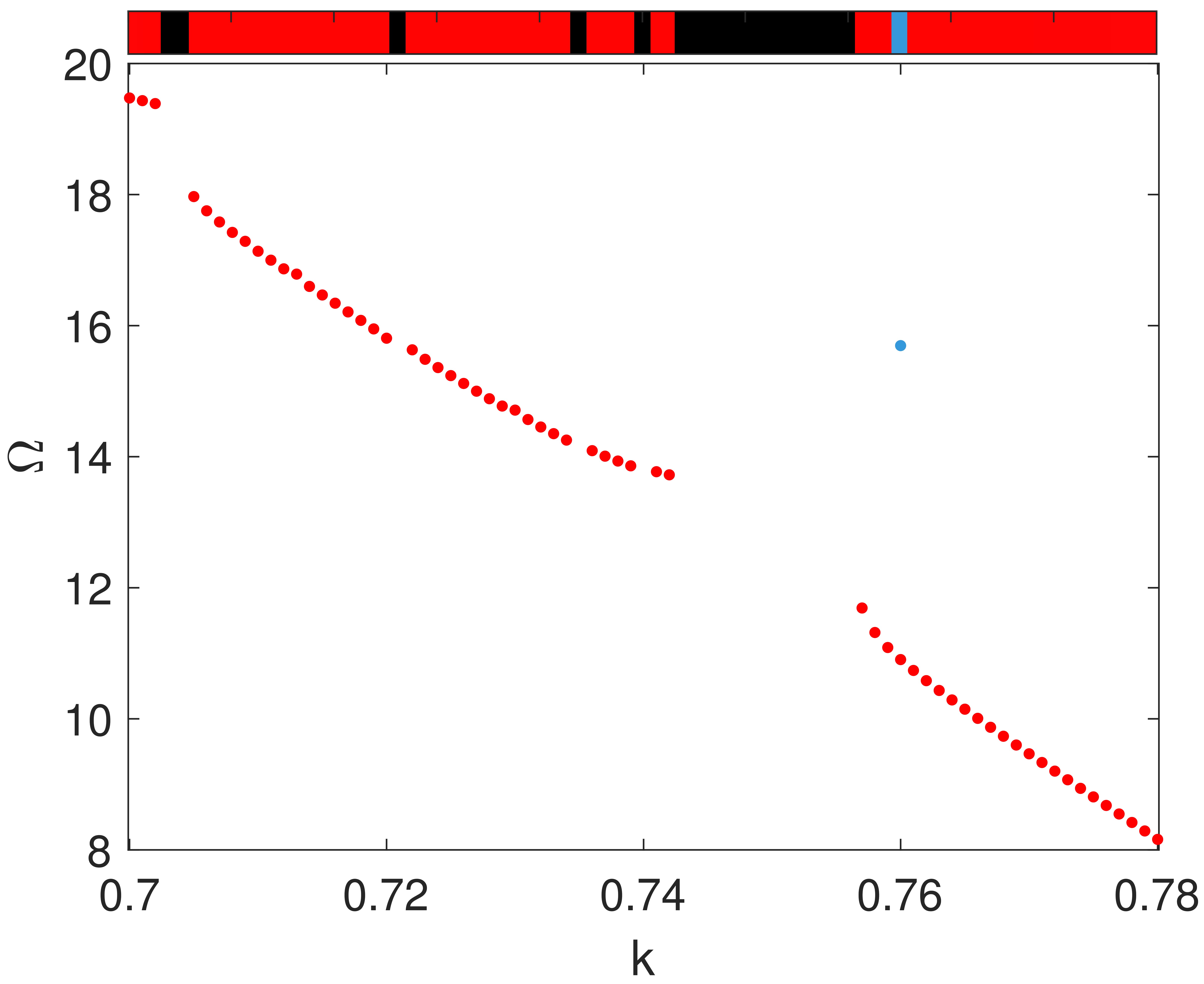}
	\caption{}
	\end{subfigure}	
  \caption{ Rotation number $\Omega$ as function of the parameter $k$. Red  (black) bars indicate the existence (non existence) of shearless barriers. The blue bar indicates a interval with two shearless barriers. (a) $\Omega$ versus $k$ for fixed $\phi_{2}=1.6\times10^{-3}$. 
(b) Zoom in the interval $0.7\leq k\leq 0.78$.} 
  \label{fig:9}
\end{figure}

To verify the influence of the radial electric field profile on the existence of the shearless barriers, we fix the fluctuation amplitude $\phi_{2}$, choosing $\phi_{2}=1.6\times10^{-3}$, and vary the parameter $k$, defined as $k= -\beta/(3\alpha)$, that corresponds to the radial position of the maximum of the electric field (normalized to the plasma radius).  The parameters $\alpha$ and $\beta$ are defined in the expression of $E_r$ introduced in equation (\ref{eq:electric_profile}) \cite{marcus}. In Sections \ref{sec:Transport-barriers} and \ref{sec:Persistent-Barriers}, $k = 0.74$. Figures \ref{fig:9}a and b show how the rotation number changes with the parameter $k$. In this figure, the red and blue bars indicate the parameter range for which barriers are identified and the black bars represent the barriers absence. Figure \ref{fig:9}b is a magnification of Figure \ref{fig:9}a.  As the radial electric field profile changes, the resonance condition also changes and this is the reason why there is no shearless barrier in the range around $k=0.74$, where the resonance condition also includes $n=2$. Figure \ref{fig:9} also shows an interval of parameter $k$ for which two barriers exist.\\

\begin{figure}[H]
  \centering
   \begin{subfigure}[b]{0.33\linewidth}
	\includegraphics[width=\linewidth]{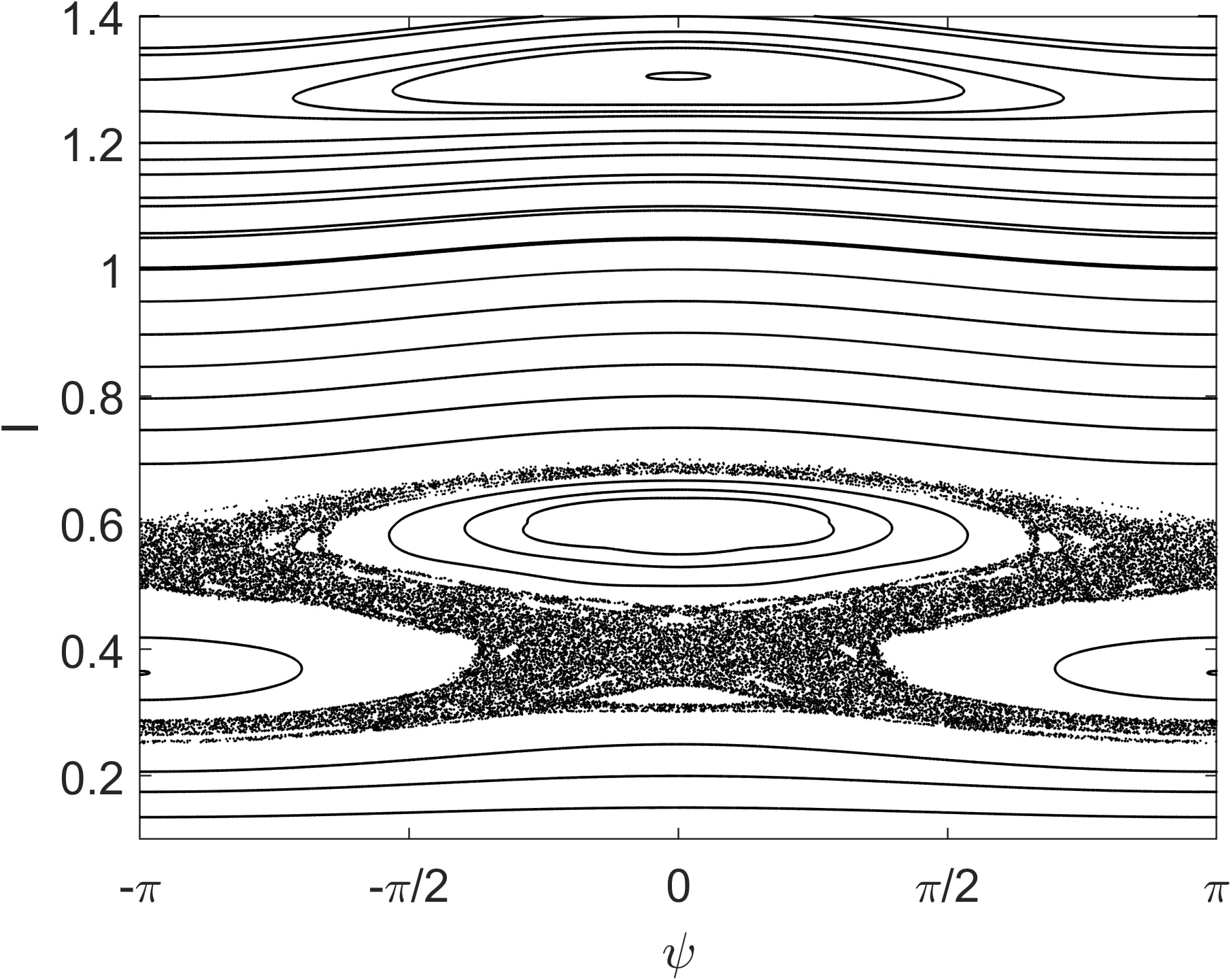}
	\caption{}
	\end{subfigure}	
	\begin{subfigure}[b]{0.33\linewidth}
	\includegraphics[width=\linewidth]{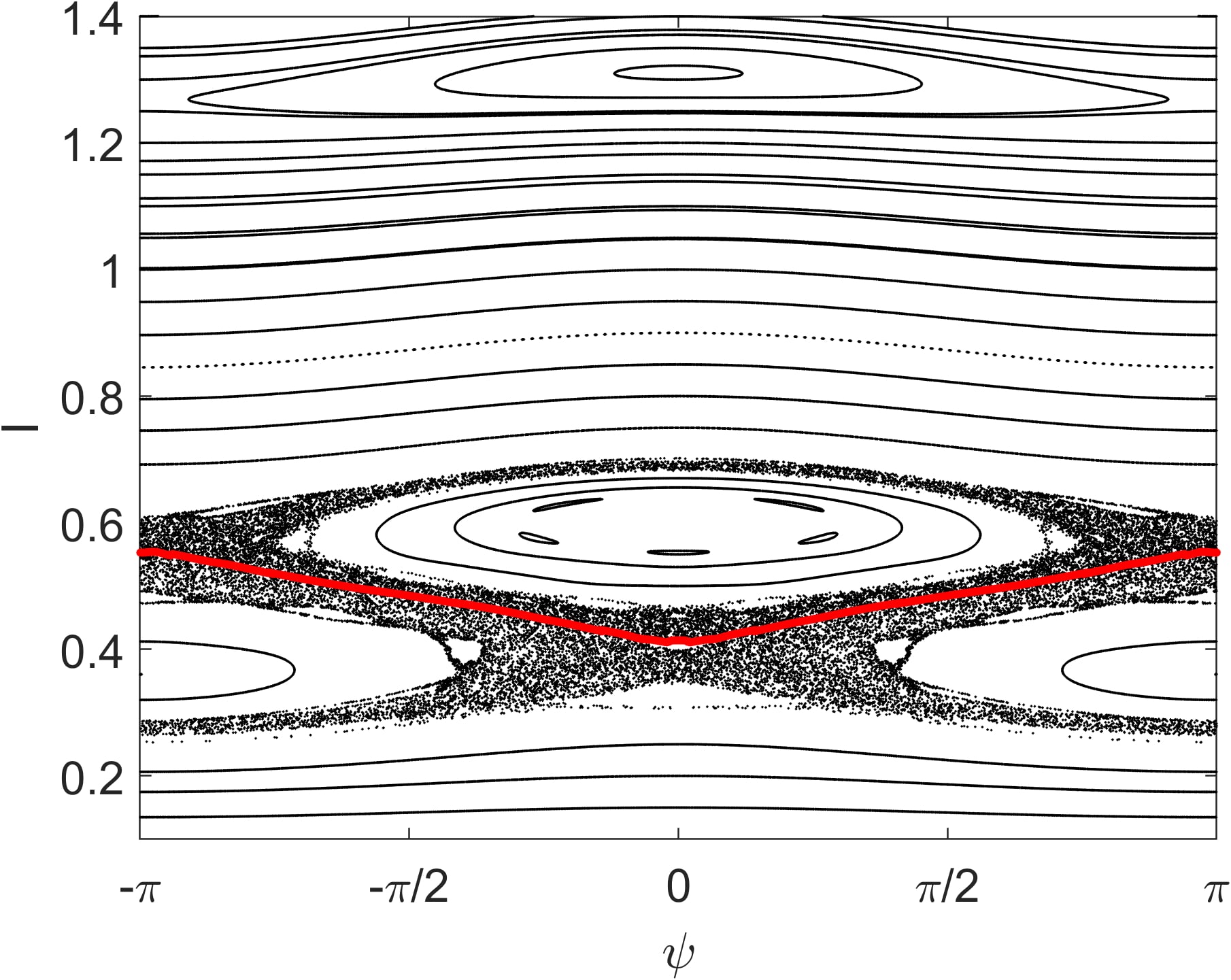}
	\caption{}
	\end{subfigure}	
  \caption{(a) No shearless barrier for k=0.704 and (b) sudden appearance of a shearless barrier around k=0.7044, marked in red.}
  \label{fig:10}
\end{figure}

It is also noted that in the interval $k=[0.7,0.78]$, as shown in the top bar of Figure \ref{fig:9}b, the shearless barrier appears and disappears on repeated occasions. To clarify this point, we show in Figure \ref{fig:10}a the Poincar{\'e} map for $k=0.704$. There is no shearless barrier in this case, only the chaotic trajectories are observed. A small modification of parameter $k$ to $k = 0.7044$, as shown in Figure \ref{fig:10}b, determines the appearance of the shearless barrier and this scenario occurs repeatedly as the electric field profile changes, as shown in Figures \ref{fig:9}a and \ref{fig:9}b.\\

\begin{figure}[H]
  \centering
  \begin{subfigure}[b]{0.33\linewidth}
	\includegraphics[width=\linewidth]{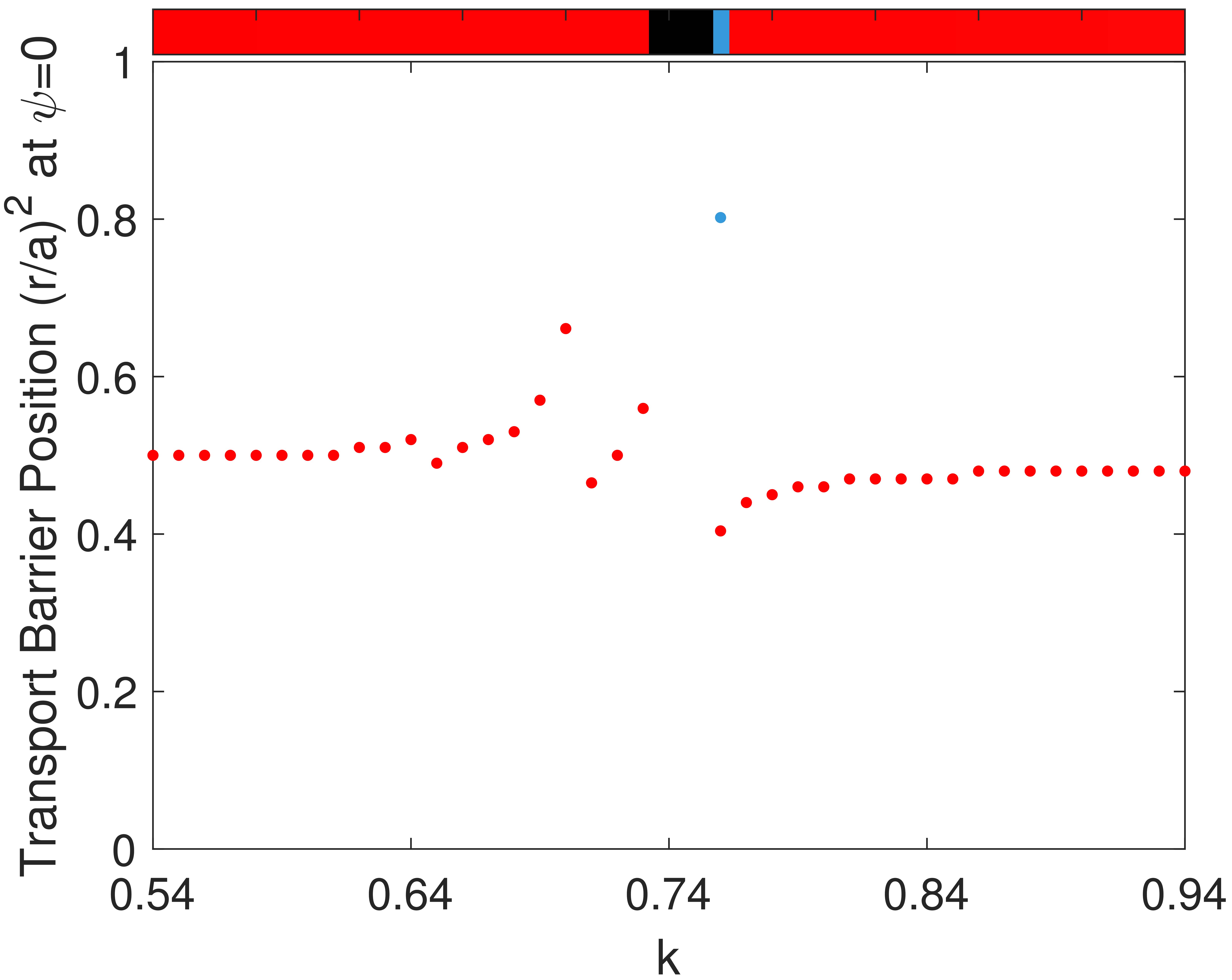}
	\caption{}
	\end{subfigure}	
	\begin{subfigure}[b]{0.33\linewidth}
	\includegraphics[width=\linewidth]{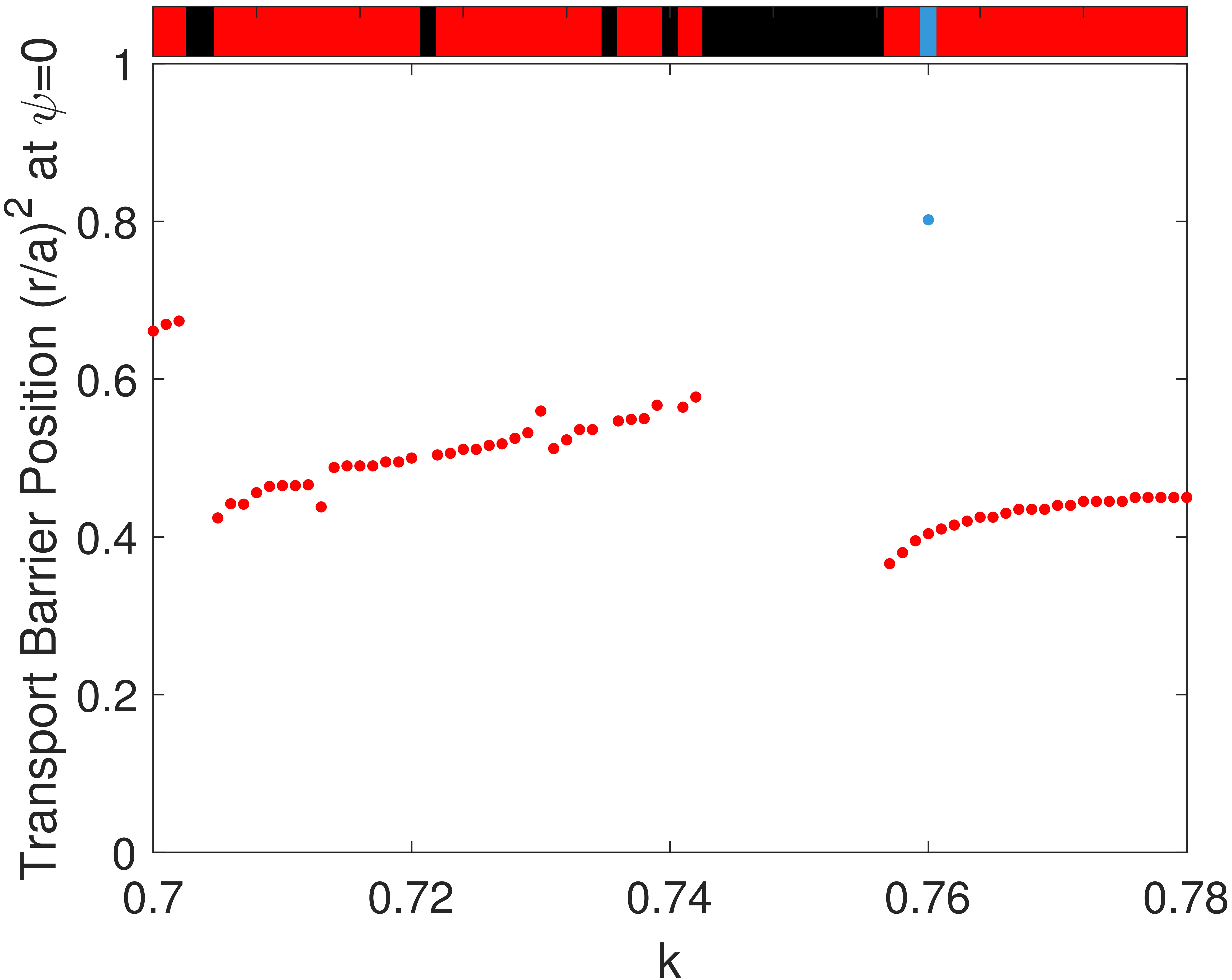}
	\caption{}
	\end{subfigure}	
  \caption{(a) Transport barrier position $I$, at $\psi=0$, versus $k$ for a fixed $\phi_2=1.6\times 10^{-3}$. (b) Zoom in the interval $0.7\leq k\leq 0.78$.} 
  \label{fig:11}
\end{figure}

Figures \ref{fig:11}a and b show the transport barrier position $I$, at $\psi=0$, as a function of $k$ for a fixed amplitude  $\phi_2=1.6\times 10^{-3}$. Figure \ref{fig:11}b is a magnification of Figure \ref{fig:11}a. Similarly to the conclusions of Figure \ref{fig:9}, on varying the parameter $k$ we find that the shearless barriers are sensitive to this parameter. Thus, a small variation of $k$, inherent to the plasma discharges evolution, can break or create such barriers or even move it towards the edge.

\section{Conclusions\label{sec:Conclusions}}

The existence of shearless transport barriers represents a novel feature in the investigation of the possible ways to control or mitigate particle transport in tokamaks and other toroidal devices like the Texas Helimak. While internal transport barriers are usually related to strong density gradients, both in the plasma edge as well as its core, shearless transport barriers appear due to a different cause, namely the existence of non-monotonic plasma profiles. These primary shearless barriers are located at the extremum points of those profiles, and do not need strong density gradients to be effective against particle transport.\\

There are three types of non-monotonic plasma profiles which can be harnessed in order to create shearless transport barriers: (i) magnetic shear (safety factor); (ii) radial electric field; and (iii) toroidal plasma velocity. In the present paper, we investigated the production of shearless transport barriers through the non-monotonic profile (ii), using a drift-kinetic model (the safety factor profile was kept monotonic). The numerical integration of the model equations is used to obtain a Poincar\'e map for canonically conjugate variables $(I,\psi)$.\\

Considering that the chaotic particle transport is influenced by a turbulent fluctuation spectrum, we considered a finite drift-wave mode spectrum for the floating electrostatic potential. The intensity of the non-resonant mode ($\phi_2$) has been used as a variable control parameter. On increasing the latter, we have shown that a primary shearless transport barrier which is destroyed for a given value of $\phi_2$ can reappear at a slightly larger value. The reason for this behavior is that shearless barriers occur at local extremum points of the rotation number, which is the average progress of the $\psi$ variable per map iteration. Alterations in the value of the perturbation strength modify the rotation number radial profiles, what can either create or destroy local extrema. Moreover, for some specific intervals of the control parameters, we identify new sequences of secondary transport barriers not associated to the non-monotonic plasma profile.  Furthermore, these secondary barriers are formed by two or three coexisting shearless invariants, constituting a noticeable obstacle to the chaotic particle transport at the plasma edge.\\

The shearless barriers described above are truly invariant curves of the Poincar\'e map of the particle trajectories. However, even after these curves are destroyed, their remnants may cause a stickiness effect which effectively traps chaotic trajectories thereby for a relatively long time. This can be also regarded as a (partial) shearless barrier. We have described those persistent barriers as the control parameter is varied.\\

The shearless barriers can also be investigated by altering the radial electric field profile, keeping constant the perturbation mode amplitude. This can be done in the context of our model by altering the position of the radial electric field extremum ($k$). As before, we can also detect the break-up and resurging of shearless barriers as the new control parameter is varied, and regions of barrier coexistence can be also observed. These multiple barriers can be related to double or triple shearless bifurcation. The effect of a small modification of the parameter k on the appearance of barriers could be validated in experiments for time-dependent electric fields.\\ 

In conclusion, varying the amplitude fluctuation  or the electric shear, we find  intervals of these control parameters
for which the barriers onset and break-up are recurrent.

\begin{acknowledgments}
The authors thank the financial support from the Brazilian Federal Agencies (CNPq), grants 407299/2018-1 and 302665/2017-0 and the S{\~a}o Paulo Research Foundation (FAPESP, Brazil) under grants 2018/03211-6, 2018/14435-2 and 2020/01399-8 and support from Coordena\c{c}\~{a}o de Aperfei\c{c}oamento de Pessoal de N{\'i}vel Superior (CAPES) under Grant Nos. 88881.143103/2017-01, Comit{\'e} Fran\c{c}ais d'Evaluation de la Coop{\'e}ration Universitaire et Scientifique avec le Br{\'e}sil (COFECUB) under Grant No. 40273QA-Ph908/18.\\
\end{acknowledgments}

DATA AVAILABILITY

The data that support the findings of this study are available
from the corresponding authors upon reasonable request.

\bibliographystyle{aipnum4-1}
\bibliography{biblio}

\end{document}